\documentclass[notitlepage,twocolumn,letterpaper,natbib,aps,prd,amsmath,amsfonts,nofootinbib,preprintnumbers,superscriptaddress,secnumarabic]{revtex4-1}
\pdfoutput=1
\usepackage{amssymb,amsmath,latexsym,mathrsfs}
\usepackage{url}
\usepackage{enumitem}
\usepackage{graphicx}
\usepackage[usenames,dvipsnames]{color}
\usepackage[breaklinks,colorlinks,urlcolor=blue,citecolor=blue,linkcolor=magenta]{hyperref}
\usepackage{multirow}
\usepackage{float}
\usepackage{cases}
\usepackage{mathtools}
\usepackage[normalem]{ulem}
\usepackage{orcidlink}
\usepackage{array}
\usepackage{soul}

\usepackage{xcolor}
\definecolor{linkcolor}{rgb}{0.0, 0.47, 0.75}
\definecolor{citecolor}{rgb}{1.0, 0.5, 0.0}
\hypersetup{
  linkcolor  = linkcolor,
  citecolor  = linkcolor,
  urlcolor   = linkcolor,
  colorlinks = true
}

\definecolor{mypink1}{rgb}{0.858, 0.188, 0.478}

\newcommand{\summnu}{\sum m_\nu}

\begin{document}

\title{Living at the Edge:\\ A Critical Look at the Cosmological Neutrino Mass Bound}

\preprint{CERN-TH-2024-115}
\preprint{IFT-UAM/CSIC-24-106}

\author{Daniel Naredo-Tuero \orcidlink{0000-0002-5161-5895}}
\email{daniel.naredo@ift.csic.es}
\affiliation{Departamento de F\'{\i}sica Te\'orica and Instituto de F\'{\i}sica Te\'orica UAM/CSIC,\\
Universidad Aut\'onoma de Madrid, Cantoblanco, 28049 Madrid, Spain}

\author{Miguel Escudero \orcidlink{0000-0002-4487-8742}}
\email{miguel.escudero@cern.ch}
\affiliation{Theoretical Physics Department, CERN, 1211 Geneva 23, Switzerland}

\author{\mbox{Enrique Fernandez-Martinez \orcidlink{0000-0002-6274-4473}}}
\email{enrique.fernandez@csic.es}
\affiliation{Departamento de F\'{\i}sica Te\'orica and Instituto de F\'{\i}sica Te\'orica UAM/CSIC,\\
Universidad Aut\'onoma de Madrid, Cantoblanco, 28049 Madrid, Spain}

\author{Xabier Marcano \orcidlink{0000-0003-0033-0504}}
\email{xabier.marcano@uam.es}
\affiliation{Departamento de F\'{\i}sica Te\'orica and Instituto de F\'{\i}sica Te\'orica UAM/CSIC,\\
Universidad Aut\'onoma de Madrid, Cantoblanco, 28049 Madrid, Spain}

\author{Vivian Poulin \orcidlink{0000-0002-9117-5257}}
\email{vivian.poulin@umontpellier.fr}
 \affiliation{Laboratoire univers et particules de Montpellier (LUPM), Centre national de la recherche scientifique (CNRS) et Universit\'e de Montpellier, Place Eug\`ene Bataillon, 34095 Montpellier C\'edex 05, France}

\date{\today}

\begin{abstract}
Cosmological neutrino mass bounds are becoming increasingly stringent. The latest limit within $\Lambda$CDM from Planck 2018+ACT lensing+DESI is $\sum m_\nu < 0.072\,{\rm eV}$ at 95\% CL, very close to the minimum possible sum of neutrino masses ($\sum m_\nu > 0.06\,{\rm eV}$), hinting at vanishing or even ``negative'' cosmological neutrino masses. 
In this context, it is urgent to carefully evaluate the origin of these cosmological constraints. In this paper, we investigate the robustness of these results in three ways:
i) we check the role of potential anomalies in Planck CMB and DESI BAO data; ii) we compare the results for frequentist and Bayesian techniques, as very close to physical boundaries subtleties in the derivation and interpretation of constraints can arise; iii) we investigate how deviations from $\Lambda$CDM, potentially alleviating these anomalies, can alter the constraints. 
From a profile likelihood analysis, we derive constraints in agreement at the $\sim 10\%$ level with Bayesian posteriors. We find that the weak preference for negative neutrino masses is mostly present for Planck 18 data, affected by the well-known `lensing anomaly'. 
It disappears when the new Planck 2020 HiLLiPoP is used, leading to significantly weaker constraints. 
Additionally, the pull towards negative masses in DESI data stems from the $z=0.7$ bin, which contains a BAO measurement in $\sim 3\sigma$ tension with Planck expectations. Without this bin, and in combination with HiLLiPoP, the bound relaxes to $\sum m_\nu < 0.11\,{\rm eV}$ at 95\% CL. The recent preference for dynamical dark energy alleviates this tension and further weakens the bound. As we are at the dawn of a neutrino mass discovery from cosmology, it will be very exciting to see if this trend is confirmed by future data. 
\end{abstract}

\maketitle
%\tableofcontents

%%%%%%%%%%%%%%%%%%%%%%%%%%%%%%%%%%%%%%%%%%%%%%%%%%%%%%%%%%%%%%%%%%%%%%%%%%%%
\section{Introduction: Neutrino Mass bounds as of mid 2024} 
%%%%%%%%%%%%%%%%%%%%%%%%%%%%%%%%%%%%%%%%%%%%%%%%%%%%%%%%%%%%%%%%%%%%%%%%%%%%
\noindent In April 2024, the DESI collaboration presented the most stringent bound on the sum of neutrino masses within the standard cosmological model~\cite{DESI:2024mwx}:
\begin{align}\label{eq:boundDESI}
    \sum m_\nu < 0.072\,{\rm eV} \,\,\,\,{[95\% \,\,{\rm CL}]}\,,
\end{align}
obtained by combining their new DESI-Y1 baryon acoustic oscillation data (BAO)~\cite{DESI:2024uvr,DESI:2024lzq} with Planck~\cite{Planck:2018vyg,Planck:2018nkj,Carron:2022eyg} and ACT data~\cite{ACT:2023kun,ACT:2023dou} (see also~\cite{Wang:2024hen,Allali:2024aiv} for further updated constraints within $\Lambda$CDM\footnote{Note that an absolute neutrino mass bound (i.e. from the laboratory and therefore independent upon the assumed cosmological model) has been recently updated by the KATRIN experiment to $\sum m_\nu < 0.93\,{\rm eV}$ at 90\% CL~\cite{Aker:2024drp} when following the Feldman-Cousins prescription. This limit is currently much weaker than the cosmological ones.}).

The bound in Eq.~\eqref{eq:boundDESI} should be compared with the minimum possible value of the sum of neutrino masses given the observed mass squared differences from neutrino oscillation experiments. Taking them from NuFITv5.3~\cite{Esteban:2020cvm} and by taking $m_{\nu, {\rm lightest}}\to 0$, one finds at 5$\sigma$ CL:
\begin{align}
    \sum m_\nu &> 0.057 \,{\rm eV} \simeq 0.06\,{\rm eV} \quad [{\rm NO}]\,, \\
    \sum m_\nu &> 0.096 \,{\rm eV} \simeq 0.10\,{\rm eV} \quad [{\rm IO}] \,,  
\end{align}
depending upon the neutrino mass ordering. At present, there is no strong preference for either ordering from global analyses of neutrino oscillation data, see~\cite{Esteban:2020cvm,deSalas:2020pgw,Capozzi:2021fjo}. 

Clearly, the current cosmological limit is very close to the minimal value in normal ordering and already disfavours to some degree the inverted one. Importantly, the limit in Eq.~\eqref{eq:boundDESI} is so close to the minimum physical boundary that statistical statements about the neutrino mass need to be taken with care. However, cosmological limits are typically derived within a Bayesian framework, and as such will depend upon the priors used. While the dependence may be weak when the likelihood largely dominates over the prior, and far away from physical boundaries, it has already been established that neutrino masses are strongly sensitive to the choice of prior~\cite{Simpson:2017qvj,Gariazzo:2018pei,Gariazzo:2022ahe,Gariazzo:2023joe}, even with DESI~\cite{DESI:2024mwx}.
Moreover, and perhaps even more surprising, there is still no hint of a non-zero mass in the posterior probability density, and in fact, it has been argued that cosmological data may favor {\it  a negative effective  neutrino mass}~\cite{Craig:2024tky,Green:2024xbb,Elbers:2024sha}. While this could be the result of a statistical fluctuation or a systematic effect, it could potentially be the indication of new phenomena in cosmology with groundbreaking implications.

Given the current situation, it is urgent to carefully analyze the origin and behaviour of the present constraints on neutrino masses from cosmology.  In this paper, we set as goals to investigate simultaneously: i) the data that have been used and the role of potential statistical anomalies in those data, ii) the statistical methods used to derive the results, and whether unwarranted effects in the Bayesian analysis drive the preference for negative neutrino masses, and iii) the extent upon which they rely on the assumption of the standard $\Lambda$CDM model.

In order to tackle these questions, we perform an extensive comparison of cosmological neutrino mass bounds derived both from a Bayesian and a frequentist standpoint. 
In particular, the use of a frequentist analysis framework can lead to new insight into the sensitivity of the bound  on the statistical procedure adopted, and can help in addressing the role of priors in the Bayesian limits. In addition, as we build the {\it likelihood profile} of the sum of neutrino masses in light of various datasets, it is possible to extrapolate to the unphysical region and study the potential preference for negative masses, as performed in previous analyses of the neutrino mass in cosmology \cite{Planck:2013nga,Couchot:2017pvz,eBOSS:2020yzd}. Our analysis is complementary to recent Bayesian method that rely on the use of an ``effective neutrino mass'' to model the effect of a negative neutrino mass \cite{Craig:2024tky,Elbers:2024sha}, and although it comes with its own set of approximations, it by-passes the need for an arbitrary definition which will necessarily miss part of the physical effect that a real model would have.

Second, to address the questions of the robustness of the bound to the choice of data, we perform series of analyses comparing results from the latest BAO data from DESI~\cite{DESI:2024uvr,DESI:2024lzq}, previous ones from SDSS~\cite{eBOSS:2020yzd}, the compilation of uncalibrated supernova (SN) distances from Pantheon+~\cite{Brout:2022vxf}, and most importantly the latest versions of the Planck likelihoods based on the 2020 PR4 data release~\cite{Planck:2020olo}. This is key because Planck currently dominates the neutrino mass bounds and because these latest likelihoods not only contain roughly ${\cal O}(10\%)$ more statistical power than the 2018 (PR3) release, but importantly because they have a much better handle of a number of systematic effects (see~\cite{Planck:2020olo} for details). Importantly, the known lensing anomaly which was present at the $2.8\sigma$ level in the Planck 2018 Plik likelihood \cite{Planck:2018vyg}  is only present in the 2020 implementations at the $1.7\sigma$ or $0.75\sigma$ levels, for the CamSpec~\cite{Efstathiou:2019mdh,Rosenberg:2022sdy} and HiLLiPoP~\cite{Tristram:2021tvh,Tristram:2023haj} likelihoods, respectively. This is well known to have an impact on the inference of neutrino masses~\cite{RoyChoudhury:2019hls,Motloch:2019gux,DiValentino:2021imh}. Yet, Eq.~\eqref{eq:boundDESI} does not make use of these newer data, and should thus be explicitly checked. In particular, we test the impact of the anomaly on the bounds in two ways: first, by simply updating the data to match the newer release (as in~\cite{Allali:2024aiv,Green:2024xbb}) and second, by explicitly marginalizing over the $A_{\rm lens}$ parameter~\cite{Calabrese:2008rt,Planck:2018vyg}, that scales the amplitude of the lensing power spectrum.

Finally, we investigate why the bound in Eq.~\eqref{eq:boundDESI} is so strong, despite the statistical power of DESI-Y1 BAO data being {\it a priori} not larger (though competitive) than that of the entire SDSS sample. We believe that this is crucial as the community expects upcoming measurements from DESI and Euclid to actually pin-down the absolute neutrino mass scale in cosmology within the next few years, see e.g.~\cite{Font-Ribera:2013rwa,EuclidTheoryWorkingGroup:2012gxx,DESI:2016fyo,Chudaykin:2019ock}. If these collaborations do not report any measurement of the neutrino mass, their analyses will clearly signal a breakdown of the standard cosmological model and possibly even new physics in the neutrino sector. This could point towards decaying neutrinos~\cite{Escudero:2020ped,Chacko:2019nej,Escudero:2019gfk,Chacko:2020hmh,FrancoAbellan:2021hdb,Barenboim:2020vrr,Chen:2022idm}, non-standard cosmic neutrino backgrounds~\cite{Farzan:2015pca,Escudero:2022gez,Alvey:2021sji,Cuoco:2005qr,Oldengott:2019lke,GAMBITCosmologyWorkgroup:2020htv}, or neutrinos with a time-varying mass~\cite{Dvali:2016uhn,Dvali:2021uvk,Lorenz:2018fzb,Lorenz:2021alz,Esteban:2021ozz,Esteban:2022rjk,Sen:2023uga,Sen:2024pgb}.
Yet, before making such claims, it is important to note that DESI data are in some (arguably small) $2\sigma$ level tension with Planck. In fact, there are two BAO measurements at $z = 0.5$ and $z=0.7$ that are driving the tension with Planck (respectively at the 2.8$\sigma$ and $2.6\sigma$ level), and that seem somewhat at odds with SDSS results as well. This raises the questions of the role of those data points in driving the strong bound in Eq.~\eqref{eq:boundDESI}. In addition, when combined with Planck and Pantheon+ (or other SN compilation) the compilation of data seem to favor dynamical dark energy over $\Lambda$CDM~\cite{DESI:2024mwx}. Such preference is known to also alter the bound given by Eq.~\eqref{eq:boundDESI}. Within our joint Bayesian and frequentist framework, we will thus explicitly check the role of those potential outlier points, as well as the preference for a deviation from $\Lambda$CDM in driving the hint of negative neutrino mass, and the strong neutrino mass bounds.

The rest of our study is structured as follows: first, in Section~\ref{sec:nuincosmo} we briefly review the main cosmological implications of massive neutrinos. We will highlight the crucial role of Planck data as well as why BAO data can significantly tighten the constraints on the neutrino mass. In Section~\ref{sec:Data} we outline the various data sets that we will use, and describe the statistical procedure (both frequentist and Bayesian) that we use to analyse the data and derive bounds. In Section~\ref{sec:results} we present our main results focusing first on analyses with Planck data only, exploring in detail the role of the lensing anomaly; we then perform comprehensive analysis of Planck and BAO data, comparing in particular DESI and SDSS; and finally, check the impact of considering extensions to $\Lambda$CDM. In Section~\ref{sec:negativeneutrinos} we specifically address the potential cosmological preference for a negative neutrino mass. Finally, in Section~\ref{sec:conclusions} we draw our conclusions.
Additional material supporting our findings, including posterior distributions, correlations and comparisons with previous works, is provided in the Appendices.

%%%%%%%%%%%%%%%%%%%%%%%%%%%%%%%%%%%%%%%%%%%%%%%%%%%%%%%%%%%%%%%%%%%%%%%%%%%%
\section{Cosmological Impact of the neutrino mass} \label{sec:nuincosmo}
%%%%%%%%%%%%%%%%%%%%%%%%%%%%%%%%%%%%%%%%%%%%%%%%%%%%%%%%%%%%%%%%%%%%%%%%%%%%

 The existence of the Cosmic Neutrino Background (C$\nu$B) is a key prediction of the Standard Model of cosmology \cite{Lesgourgues:2013sjj}. The C$\nu$B would have formed in the early Universe at temperatures of $\sim 2\,{\rm MeV}$~\cite{Dolgov:2002wy}, and according to the $\Lambda$CDM model, we should be living in a Universe filled with a number density of neutrinos of $n_\nu \simeq 56/{\rm cm}^3$ per helicity state. These neutrinos gravitate today primarily as a result of their mass and in fact their energy density today is not negligible:
\begin{align}\label{eq:Omeganuh2}
  \Omega_\nu h^2 = \frac{\sum m_\nu n_\nu}{\rho_{\rm crit}} \simeq \frac{\sum m_\nu}{93.2\,{\rm eV}} \simeq 0.0012\frac{\sum m_\nu}{0.12\,{\rm eV}}\,.
\end{align}
This is $\mathcal{O}(1)\%$ of the dark matter energy density.

The implications of the neutrino mass in cosmology have been explored and discussed in depth and length, see~\cite{Lesgourgues:2013sjj,Dolgov:2002wy,Lesgourgues:2006nd,Hannestad:2010kz,Wong:2011ip,Lattanzi:2017ubx} for reviews. In essence, the cosmological implications of neutrino masses can be understood as follows: 1) Neutrinos are always a relevant component of the energy density of the Universe and therefore contribute to its expansion rate, $H\propto \sqrt{\rho}$. 2) Neutrinos were propagating ultrarelativistically until the Universe cooled down to $T_\nu \simeq m_\nu/3$ which occurs at $z_{\nu}^{\rm NR} \simeq 190\, m_\nu/(0.1\,{\rm eV})$. 3) As a result of their ultrarelativistic speeds, neutrinos are not able to cluster on scales smaller than $L\simeq 20\,{\rm Mpc}\, 0.1\,{\rm eV}/m_\nu$ today. % See Eq. 8.83 of Dodelson
In consequence, and for massive neutrinos which become non-relativistic after recombination $m_\nu \lesssim 0.6\,{\rm eV}$, one of their main cosmological implications is to suppress the amount of structure formation on scales smaller than $L\simeq 20\,{\rm Mpc}\, 0.1\,{\rm eV}/m_\nu$. This suppression is of course strongly dependent upon the energy density neutrinos represent in the Universe which is directly proportional to the neutrino masses as given in Eq.~\eqref{eq:Omeganuh2}. This discussion clearly highlights that the best way to search for the neutrino mass is arguably by directly observing probes of the Large Scale Structure in the Universe. This is one of the primary goals of the ongoing DESI and Euclid surveys 
(see~\cite{Font-Ribera:2013rwa,EuclidTheoryWorkingGroup:2012gxx,DESI:2016fyo}). Nevertheless, as of today the bound on the neutrino mass is dominated by Planck with the aid of BAO and Supernova measurements to break relevant parameter degeneracies. 

\begin{figure}[t!]
    \centering
    \includegraphics[width=0.48\textwidth]{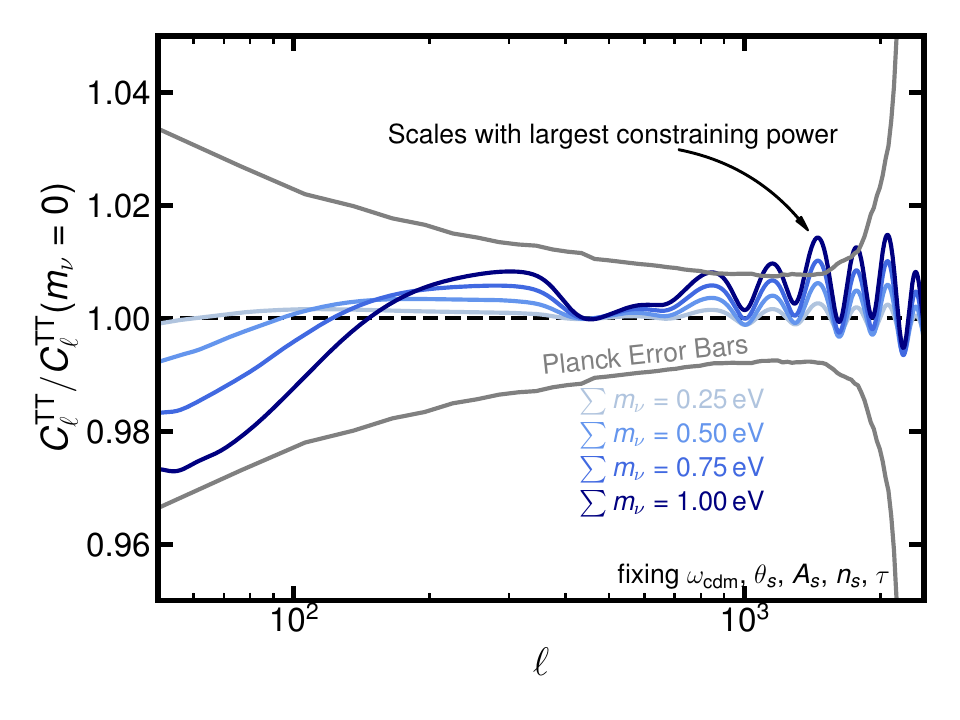}
    \caption{Impact of a non-zero neutrino mass on the TT power spectrum. Inspired by Figure 26.2 of~\cite{Workman:2022ynf} but showing the Planck error bars taken from the binned PR3 data release\footnote{\url{https://irsa.ipac.caltech.edu/data/Planck/release_3/ancillary-data/}}. Note that it is precisely in the range of angular scales where the main impact of neutrino masses appears where the lensing anomaly is present in some Planck likelihood implementations.}
    \label{fig:CLTTPlanck}
\end{figure}

How is then Planck sensitive to the neutrino mass if neutrinos with $m_\nu < 0.6\,{\rm eV}$ became non-relativisitic after the CMB was formed? There are two effects that matter~\cite{Archidiacono:2016lnv}: Firstly, as a result of neutrinos traveling at relativistic speeds, they lead to a suppression of the matter power spectrum. This reduces the lensing that CMB photons experience on their way from the last scattering surface until today, and this in turn leads to sharper peaks in the CMB power spectrum. This effect is depicted in Fig.~\ref{fig:CLTTPlanck}. In this figure, several other cosmological parameters have been fixed: the total matter densities of baryons ($\omega_b$) and dark matter $(\omega_{\rm cdm})$, the Thomson optical width to reionization $\tau$, the shape and amplitude of the primordial matter power spectrum ($n_s$ and $A_s$), as well as the angular scale of the first CMB peak, $\theta_s$. From Fig.~\ref{fig:CLTTPlanck} we can clearly see that the main constraining power for neutrino masses from Planck will come from rather small angular scales. Importantly, this lensing effect is also dependent upon what the values of $A_s$ and $\omega_m$ are. 

Secondly, $\sum m_\nu$ impacts the angular diameter distance to the last-scattering surface as neutrinos contribute to the Universe's expansion. It is however possible to exploit the geometrical degeneracy in the CMB by adjusting $H_0$ to compensate the effect of $\sum m_\nu$ on the last scattering surface and ensure the angular diameter distance to the last-scattering surface is left unaffected. This is done in Fig.~\ref{fig:CLTTPlanck}, by keeping $\theta_s$ fixed, and requires a smaller $H_0$. However, in this process $\Omega_m \equiv \omega_m/h^2$ will increase as a result of the necessary decrease in the reduced Hubble parameter $h\equiv H_0/100$ km/s/Mpc. This suggests that an accurate probe of the late-time expansion history can help break the geometrical degeneracy (namely $\Omega_m$ and/or $H_0$) and further constrain $\sum m_\nu$.

Therefore, in order to make precise inferences on the neutrino mass from CMB observations one needs to have i) a very good understanding of how the CMB is being lensed on small angular scales, ii) control over $A_s$ and $\omega_m$, and iii) control over $\Omega_m$ and more generally the expansion history at late-times. 

In this context, regarding i), the Planck legacy analysis did report the so-called ``lensing anomaly'' which can impact the inferences on the neutrino masses. The anomaly was parametrized introducing an \emph{ad hoc} variable $A_{\rm lens}$ to change the lensing power relative to its actual physical value, so as to account for possible systematic uncertainties, and Planck data preferred values of $A_{\rm lens}>1$. However, the magnitude of the $A_{\rm lens}$ was known to vary depending on the specific likelihood (and specific CMB dataset) used to perform the analysis. While the anomaly is $2.8\sigma$ with the official Planck collaboration 2018 TTTEEE likelihood, dubbed \texttt{plik}, it is reduced after the final PR4 data release to the 1.7$\sigma$ level with the alternative CamSpec~\cite{Efstathiou:2019mdh,Rosenberg:2022sdy} likelihood, and down to 0.75$\sigma$ with the HiLLiPoP~\cite{Tristram:2021tvh,Tristram:2023haj} one. 

Importantly, the CamSpec and HiLLiPoP likelihoods have been recently updated since the 2018 analysis, in light of a new set of maps produced by the Planck collaboration called NPIPE. The NPIPE maps exploit a number of improvements in the processing of time ordered data to allow for an increase in the signal-to-noise ratio at small scales. They also allow to use a larger sky fraction, and incorporate a better handling of a number of systematic errors thanks to dedicated mock data. This results in a roughly $\sim 10\%$ stronger constraining power on $\Lambda$CDM parameters, and importantly, the lensing anomaly seems to be significantly reduced or even absent in those data. In fact, it has been shown that they lead to {\it weaker} bounds to $\sum m_\nu$ than the Plik 2018 likelihood, see~\cite{Rosenberg:2022sdy,Tristram:2023haj,Allali:2024aiv}. 

\begin{figure}[t]
    \centering
    \includegraphics[width=0.48\textwidth]{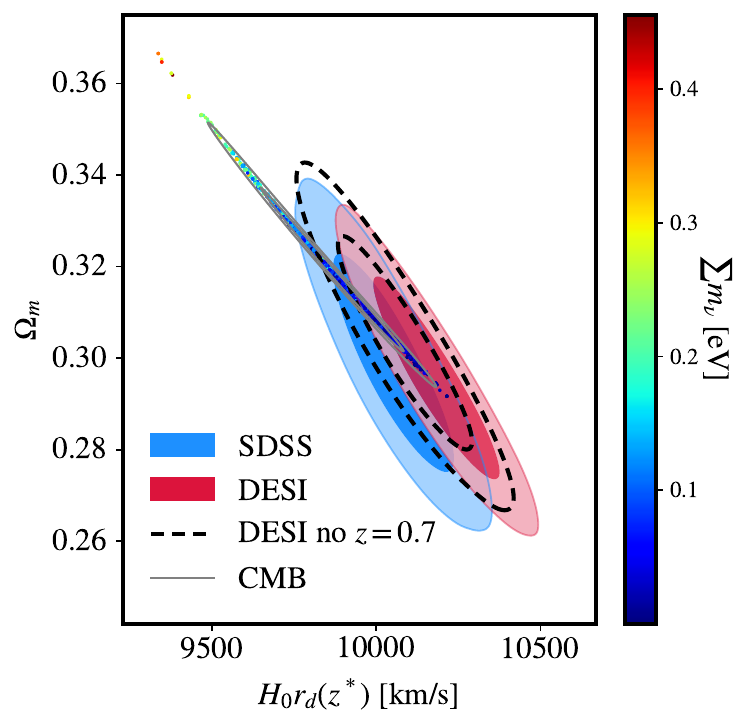}
    \caption{Implications of BAO measurements of $\Omega_m$ and $H_0 r_d$ for $\sum m_\nu$ inferences. We show the posterior density contours using Planck data (grey and dots), as well as the regions favoured by the full SDSS BAO sample (in blue), DESI-Y1 (in red), and DESI-Y1 without the $z=0.7$ bin, which contains a $2.6\sigma$ outlier (in black dashed). We can clearly see that the $z=0.7$ leads to a relevant shift on the parameter space with implications for the neutrino mass.}
    \label{fig:mnu_Omegam_vs_H0rd}
\end{figure}

Regarding ii), the best way to obtain better measurements on $A_s$ is from large scale CMB polarization measurements which will be provided by LiteBIRD~\cite{LiteBIRD:2022cnt} but on a $\sim \! 10$ year time-scale. Importantly, regarding iii), the improvement is happening now, as DESI is taking data and has published already the 1st year data release, while Euclid is on space and will start collecting data in one year as well. In this regard, it is important to highlight that the compatibility between the new DESI-Y1 data and Planck is at the $2\sigma$ level, and is thus worse than for SDSS. As a result, this (arguably small) tension may impact the neutrino mass bound. This is illustrated in Fig.~\ref{fig:mnu_Omegam_vs_H0rd}, where we compare constraints under $\Lambda$CDM in the $\Omega_m-H_0r_d$ plane from SDSS, DESI and DESI without the data points at $z=0.7$. We also superimpose the posteriors from a fit to Planck 2018 (with lensing), highlighting the correlations with $\sum m_\nu$ with colored points. As argued in Ref.~\cite{DESI:2024mwx}, it is clear that the mismatch in $H_0r_d$ between SDSS and DESI is driving the discrepancy, and a stronger bound to $\sum m_\nu$. However, removing the data points at $z=0.7$ can significantly shift the mean of the posterior distributions (by about $\sim 0.8\sigma$) without affecting the error bars. This suggests than an analysis without these data points may lead to significantly weaker bounds, and would argue in favor of a potential statistical fluke driving these bounds. 

In summary, it is very important to understand how relevant are possible systematic effects in Planck CMB data on our inferences of the neutrino mass in cosmology, as well as what are the implications of adding different sets of BAO data. This is particularly the case given that the direct combination of Planck + DESI-Y1 BAO yields a bound $\sum m_\nu < 0.072\,{\rm eV}$ which is very close to the minimum allowed value from neutrino oscillation experiments $\sum m_\nu > 0.06\,{\rm eV}$.

%%%%%%%%%%%%%%%%%%%%%%%%%%%%%%%%%%%%%%%%%%%%%%%%%%%%%%%%%%%%%%%%%%%%%%%%%%%%
\section{Data and Methodology}\label{sec:Data}
%%%%%%%%%%%%%%%%%%%%%%%%%%%%%%%%%%%%%%%%%%%%%%%%%%%%%%%%%%%%%%%%%%%%%%%%%%%%
\subsection{Cosmological Data: CMB, BAO and Supernova}
In what follows, we will first perform a comprehensive analysis of the bound on neutrino masses coming from considering Planck data alone, to highlight the role of potential anomalies (whether a statistical fluke or a systematic effect) in the data, and how subsequent data releases have affected those bounds. To that end, we consider the following likelihood combinations:
\begin{itemize}
%    \item \texttt{Planck13-PR1} -- We consider the low $\ell$ and high $\ell$ TT from the Planck 2013 data release~\cite{Planck:2013pxb}, in combination (as it was done at the time) with large scale polarization from WMAP-9~\cite{WMAP:2012fli}. 
%    \item \texttt{Planck15-PR2} --  We consider the whole low $\ell$ and high $\ell$ TTTEEE data from the Planck 2015 data release~\cite{Planck:2015fie}.
    \item \texttt{Planck18-PR3} -- We consider the default \texttt{plik} Planck legacy likelihoods for both TT, TE, EE  high $\ell$ spectra as well as the large scale (low $\ell$) EE polarization likelihood \texttt{SimAll}, and also the large scale TT \texttt{Commander} likelihood. 
    \item \texttt{CamSpec22-PR4} -- We consider the new \texttt{Planck\_CamSpec\_NPIPE12\_7\_TTTEEE} likelihood for both TT, TE, EE  high $\ell$ spectra~\cite{Efstathiou:2019mdh,Rosenberg:2022sdy} as well as the large scale (low $\ell$) EE polarization likelihood \texttt{SimAll} and the large scale TT \texttt{Commander} one. 
    \item \texttt{HiLLiPoP23-PR4} -- We consider the new \texttt{HiLLiPoP}~\cite{Tristram:2023haj} likelihood for both TT, TE, EE  high $\ell$ spectra, and the \texttt{LoLLiPoP}~\cite{Tristram:2020wbi,Tristram:2021tvh} EE, EB and BB low $\ell$ spectra.
    \item \texttt{Lensing-PR3} -- Unless otherwise stated (no lensing) we also consider the reconstructed gravitational lensing potential power spectra from Planck PR3~\cite{Planck:2018nkj}\footnote{We stress that the intrinsic lensing induced smoothing of the CMB peaks is always included.}. We note, however, that there is a newer PR4 one~\cite{Carron:2022eyg}, as well as a complementary one by ACT~\cite{ACT:2023kun,ACT:2023dou}. Including these data sets will not alter our conclusions and will only improve the bound by $\sim 10\%$ (compare Eq.~\eqref{eq:boundDESI} with Eq.~\eqref{eq:mnuBoundPlanckDESIBayesian}). 
\end{itemize}

Secondly, we will investigate the role of BAO data, and the impact of potential statistical fluctuations within DESI data on the bounds. We include the following data combinations:
\begin{itemize}
    \item \texttt{DESI-Y1} -- We consider the full set of BAO data as reported in~\cite{DESI:2024uvr,DESI:2024mwx}. 
    \item \texttt{DESI-Y1-no07} -- We consider the full set of BAO data as reported in~\cite{DESI:2024uvr,DESI:2024mwx} but without including the two data points at $z =0.7$. Since this corresponds precisely to the redshift of dark energy domination and this bin contains a data point in $\sim \! 3\sigma$ tension\footnote{Out of the 22 BAO data points, most of them are in good agreement with Planck $\Lambda$CDM predictions, but there are two DESI-Y1 data points at $z = 0.5$ and $z = 0.7$ which deviate by $\sim 2.8\sigma$ and $\sim 2.6\sigma$, respectively. Given a total of 22 total measurements of BAO, the chances of this occurring (assuming Gaussian and uncorrelated errors) is of only $\sim 1.8\%$, see~\cite{Efstathiou:2024dvn}. We further note that the DESI-Y1 BAO measurements at $z=0.7$ are also in some tension with the old SDSS ones at similar redshifts.} with Planck predictions, it is interesting to explore its impact on the $\sum m_\nu$ bound to understand what would happen if these outliers are not confirmed by future data, as highlighted in Figure~\ref{fig:mnu_Omegam_vs_H0rd}.
    \item \texttt{SDSS-full} -- We consider the full set of BAO data from SDSS as detailed in~\cite{eBOSS:2020yzd}. 
    \item \texttt{DESI-Y1+SDSS} -- We consider the DESI and SDSS combination as done in~\cite{DESI:2024mwx} that makes use of the BAO measurements from the survey with the largest effective volume in a given redshift bin. Note that this includes the DESI-Y1 BAO measurements at $z = 0.7$ which are in $\sim 3\sigma$ tension with Planck $\Lambda$CDM predictions.
\end{itemize}

Finally, we also consider analyses including uncalibrated luminosity distance-redshift measurements from type Ia SN:
\begin{itemize}
    \item \texttt{SN-Pantheon} -- We make use of the Pantheon+ catalog of uncalibrated luminosity distance of type Ia supernovae (SNeIa) in the range ${0.01<z<2.3}$~\cite{Brout:2022vxf}. We note that there are alternative compilations such as the Union3 \cite{Rubin:2023ovl} and DES-Y5 SNeIa \cite{DES:2024tys} that could be used. We do not anticipate them to strongly impact our conclusions. We leave a dedicated analysis to future work.
\end{itemize}

%%%%%%%%%%%%%%%%%%%%%%%%%%%%%%%%%%%%%%%%%%%%%%%%%%%%%%%%%%%%%%%%%%%%%%%%%%%%
\subsection{Analysis methodology: comparing Bayesian and frequentist framework}\label{sec:Methods}
%%%%%%%%%%%%%%%%%%%%%%%%%%%%%%%%%%%%%%%%%%%%%%%%%%%%%%%%%%%%%%%%%%%%%%%%%%%%

In the cosmology community, it has become standard to perform analyses through a Bayesian framework, as those are typically numerically less expensive than frequentists analyses given the very large number of (cosmological and nuisance) parameters that must be considered. These also have the claimed advantage to incorporate prior knowledge (or lack thereof) in a straightforward manner, as {\it priors} are rooted in the definition of the {\it posterior} distribution within Bayes theoreom. In addition, there are now advanced tools (based on the notion of Bayesian evidence) to perform model comparison that are routinely used in Cosmology, and can (somewhat) easily tackle the problem known as ``look elsewhere effect'' in the frequentist framework, and quantify the abstract notion of ``Occam's razor'' that is often put forward. Yet, given that cosmology bounds on the sum of neutrino masses are pushing against the physical boundary,  they can be affected by {\it prior effects:} the credible intervals built from the Bayesian posteriors become largely influenced by the choice of priors, rather than the data likelihood. As the choice of priors carry a level of arbitrariness, these effects can lead to constraints that are not robust. This is particularly relevant given the current context, and the apparent strength of the cosmological bound.

Indeed, over the past years, almost the entirety of each subsequent data release increased progressively the existing bound, with a preferred value at $\sum m_\nu = 0$ and barely any hint for a non-vanishing value, which is surprising and worth investigating. In fact, present constraints have almost exhausted the available parameter space given the lower bound from neutrino oscillations and are already in significant tension with the minimal value implied by an inverted ordering, still allowed by oscillation results. In this context, subtleties when deriving the limits must be taken into consideration. 
In a Bayesian framework, the results are therefore prior-dependent (see e.g.~Ref.~\cite{DESI:2024mwx}), as the preferred value appears to always be the smallest possible one given the prior (ie., $\sum m_\nu = 0$, 0.06 or 0.1~eV when disregarding the results from oscillations or assuming a normal or inverted ordering, respectively) and thus (artificially) relax the larger the minimal allowed mass is.

As an alternative to the Bayesian approach, and to compare and contrast the results, we make use of the frequentist framework, relying on a series of $\chi^2$ optimizations at fixed neutrino mass, to build a profile likelihood and derive confidence intervals. The main advantages of the frequentist approach are that i) constraints are insensitive to the specific choice of priors, and ii) the presence of the physical boundary can be accounted for in a statistically consistent manner~\cite{Feldman:1997qc}. Thus, a direct comparison between the two approaches may provide a better handle on the relevance of these subtleties, as well as on the robustness of the cosmological bound.

All power spectra in our study are obtained from the Boltzmann code CLASS~\cite{Blas:2011rf}. The main physical quantity impacting cosmological observations is the total energy density in non-relativistic neutrinos and as such, for simplicity, we model the neutrinos as fully degenerate with a mass $m_\nu = \sum m_{\nu}/3$, and unless specified, vary it within the prior range $m_\nu\in[0,1]$ eV. Cosmological data cannot be sensitive to the mass splittings and this choice can only cause relatively small changes in $\Delta \chi^2$, see e.g.~\cite{Couchot:2017pvz}. We additionally vary the following six cosmological parameters, within large flat priors (when applicable): the angular size of the sound horizon $\theta_s$, the physical baryon $\omega_b$ and dark matter $\omega_{\rm cdm}$ densities, the amplitude $A_s$ and tilt $n_s$ of the primordial power spectrum for scalar modes, the optical depth to reionization $\tau_{\rm reio}$.

On the Bayesian side, we perform a Markov Chain Monte Carlo (MCMC) sampling of the posterior distribution, using the publicly available \texttt{MontePython} code~\cite{Brinckmann:2018cvx,Audren:2012wb}. Cosmological and nuisance parameters are varied according to the ``fast'' and ``slow'' parameters decomposition \cite{Lewis:2013hha}.  The chains are then marginalized with \texttt{GetDist}~\cite{Lewis:2019xzd} in order to extract the bounds. On the frequentist side, instead of posterior marginalization, the relevant procedure is likelihood profiling, which requires the minimization of the $\chi^2$ function for a fixed value of the parameter of interest ($\sum m_\nu$ in our case), varying simultaneously the $N-1$ remaining cosmological and nuisance parameters. In particular, the various likelihood combinations described in Section~\ref{sec:Data} contain of the order of $N\sim 20-25$ parameters making the minimization a highly non-trivial task. In this context, we have opted to perform the numerical minimization with the simulated annealing algorithm implemented in \texttt{Procoli}~\cite{Karwal:2024qpt}\footnote{We note that there are other public codes available such as \texttt{Prospect}~\cite{Holm:2023uwa} and \texttt{CAMEL}~\cite{Henrot-Versille:2016htt}. While \texttt{Prospect} also relies on a simulated annealing algorithm and would likely provide similar results as \texttt{Procoli}, \texttt{CAMEL} relies on the quasi-Newtonian optimizer \texttt{Minuit}~\cite{James:1975dr}. We have not been able to obtain converged likelihood profiles with \texttt{CAMEL}, highlighting one of the difficulties in the frequentist approach.}. Although this procedure is straightforward, the HiLLiPoP likelihood requires significantly more care to successfully converge. We attribute this to the fact that HiLLiPoP is the sole CMB likelihood that is not binned, and is therefore noisier than the others. For this reason, our profiling is not performed over a regular grid of $\sum m_\nu$. This is, however, not a major cause for concerns, since to properly calibrate the frequentist test statistics we will in any case fit the likelihood profiles.
In addition, we can extrapolate the resulting fits to the unphysical negative neutrino mass region, allowing us to asses the potential preference for negative masses recently displayed by some datasets and discussed in~\cite{Craig:2024tky,Green:2024xbb,Elbers:2024sha}.

Consequently, in order to gauge the impact of approaching the physical boundary at $\summnu=0$, we will choose two different prescriptions to extract the frequentist bound:
\begin{enumerate}
    \item Bounded Likelihood (B.L.): focusing only on the physical region $\summnu>0$, we derive the bound via the standard $\Delta\chi^2$ cut assuming Wilks' theorem, see e.g.~\cite{Rolke:2004mj}. 
    
    \item Feldman-Cousins (F.C.): we fit the $\chi^2$ profile in the physical region to a parabola and extrapolate into the unphysical region so as to find where the true minimum would lie. It is then possible to extract the corrected upper bound from Table X of Ref.~\cite{Feldman:1997qc}. Compared with the previous procedure, this prescription has the advantage of guaranteeing proper coverage of the interval. However, it relies on the extrapolation in order to find the position and depth of the minimum. 
\end{enumerate}

In order to assess whether a $\chi^2$ profile is Gaussian and the Feldman-Cousins method can be safely applied, we will perform a parabolic fit of all points that lie below a certain $\Delta\chi^2_{\text{max}}$. By varying this parameter in the interval $\Delta\chi^2_\text{max}\in [2,4]$, we can have a measure of how much the best-fit parabola or, equivalently, the F.C. bound, depends on the amount of points that are being fit. Following this method, we will only quote frequentist bounds on the profiles that exhibit stable best-fit parabolas. We find that this is the case for all of the analyses and that the limits derived from the Feldman-Cousins procedure do not appreciatively depend upon $\Delta\chi^2_\text{max}$.

%%%%%%%%%%%%%%%%%%%%%%%%%%%%%%%%%%%%%%%%%%%%%%%%%%%%%%%%%%%%%%%%%%%%%%%%%%%%
\section{Cosmological neutrino mass bounds: A frequentist vs Bayesian comparison}\label{sec:results}
%%%%%%%%%%%%%%%%%%%%%%%%%%%%%%%%%%%%%%%%%%%%%%%%%%%%%%%%%%%%%%%%%%%%%%%%%%%%

\begin{figure*}[t]
    \centering
    \includegraphics[width=0.8\textwidth]{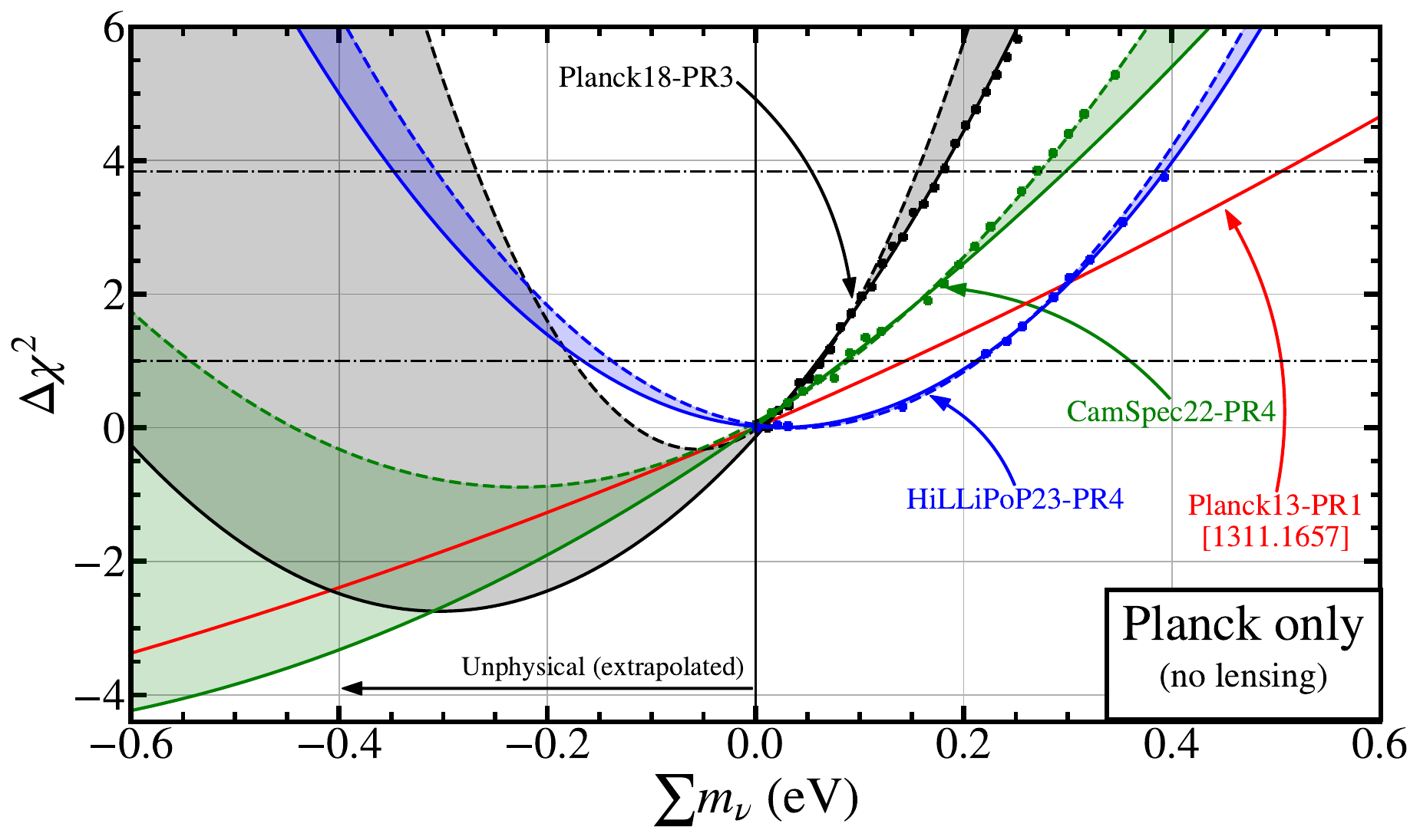}
    \caption{Profile likelihoods for the neutrino masses within $\Lambda$CDM for three different versions of the Planck likelihoods: \texttt{Planck18-PR3}, \texttt{CamSpec22-PR4}, and \texttt{HiLLiPoP23-PR4}. 
    For comparison purposes we also show the Planck 2013 results from~\cite{Planck:2013nga} in red where the potential trend for a best fit in the ``negative" regime was first highlighted. 
    We clearly see that the bound on the neutrino masses changes significantly for each implementation of the likelihood, being HiLLiPoP the one giving the looser constraints. Solid and dashed lines correspond to parabolic fits where the $\Delta \chi^{2}$ points up to 4 or 2 were used in the fit, respectively.}
    \label{fig:mnu_planck_only}
\end{figure*}

\subsection{Planck Only Analyses}\label{sec:planckonly}

\noindent \emph{Planck Legacy vs Planck 2020 likelihoods} -- In Fig.~\ref{fig:mnu_planck_only} we compare the profile likelihoods for the neutrino mass using the latest implementations of the Planck likelihoods. We also show the first Planck 2013 results from~\cite{Planck:2013nga} in red that  already highlighted the potential preference for ``negative" neutrino masses. We show results from the 2018 legacy PR3 (black) as well as the latest implementations of the PR4 data release using CamSpec (green) and HiLLiPoP (blue). As discussed in the introduction, these implementations feature different levels of lensing discrepancies which are critical for neutrino mass inferences.  The lensing anomaly in the \texttt{Planck18-PR3} one is $2.8\sigma$, in \texttt{CamSpec22-PR4} it is $1.7\sigma$, while in \texttt{HiLLiPoP23-PR4} it is only $0.75\sigma$. From Fig.~\ref{fig:mnu_planck_only}, one can see that the strength of the neutrino mass bound directly anti-correlates with the level of the anomaly, and can relax by up to a factor of 2 when going from \texttt{Planck18-PR3} to \texttt{HiLLiPoP23-PR4}. One can also clearly notice that the extrapolation to the unphysical region indicates preference for negative neutrino masses for \texttt{Planck18-PR3} and \texttt{CamSpec22-PR4},  the two likelihoods that carry some residual lensing anomaly. However, for \texttt{HiLLiPoP23-PR4}, which has no statistically significant $A_{\rm lens}$ anomaly, the minimum of the $\chi^2$ is consistent with positive (albeit small) neutrino masses.

In Table~\ref{tab:Planckonly} we explicitly show the 95\% CL bounds on the neutrino mass for the various likelihoods. Given the impact of the lensing anomaly in plik and CamSpec, their preference for a best-fit in the negative mass region is very strong. As such, our samples in the physical region lie very far from the $\Delta \chi^2$ minimum where the Gaussian approximation holds and the extrapolation has extremely large uncertainties as shown in Fig.~\ref{fig:mnu_planck_only}. Therefore, the simple Feldman-Cousins prescription when a Gaussian behaviour is observed cannot be implemented in these two cases. Instead, a full boostraping of the parameter space to calibrate the test statistic and correctly asses at which values of the $\Delta \chi^2$ lay the cuts for the confidence levels of interest would be needed. This procedure would be extremely computing-expensive and is unfortunately not feasible. On the other hand, for the HiLLiPoP implementation (which does not feature a lensing anomaly), solid frequentits limits can be obtained. A direct comparison shows that this frequentists and Bayesian bounds agree within 20\%.

\begin{figure}[t]
    \centering
    \includegraphics[width=0.48\textwidth]{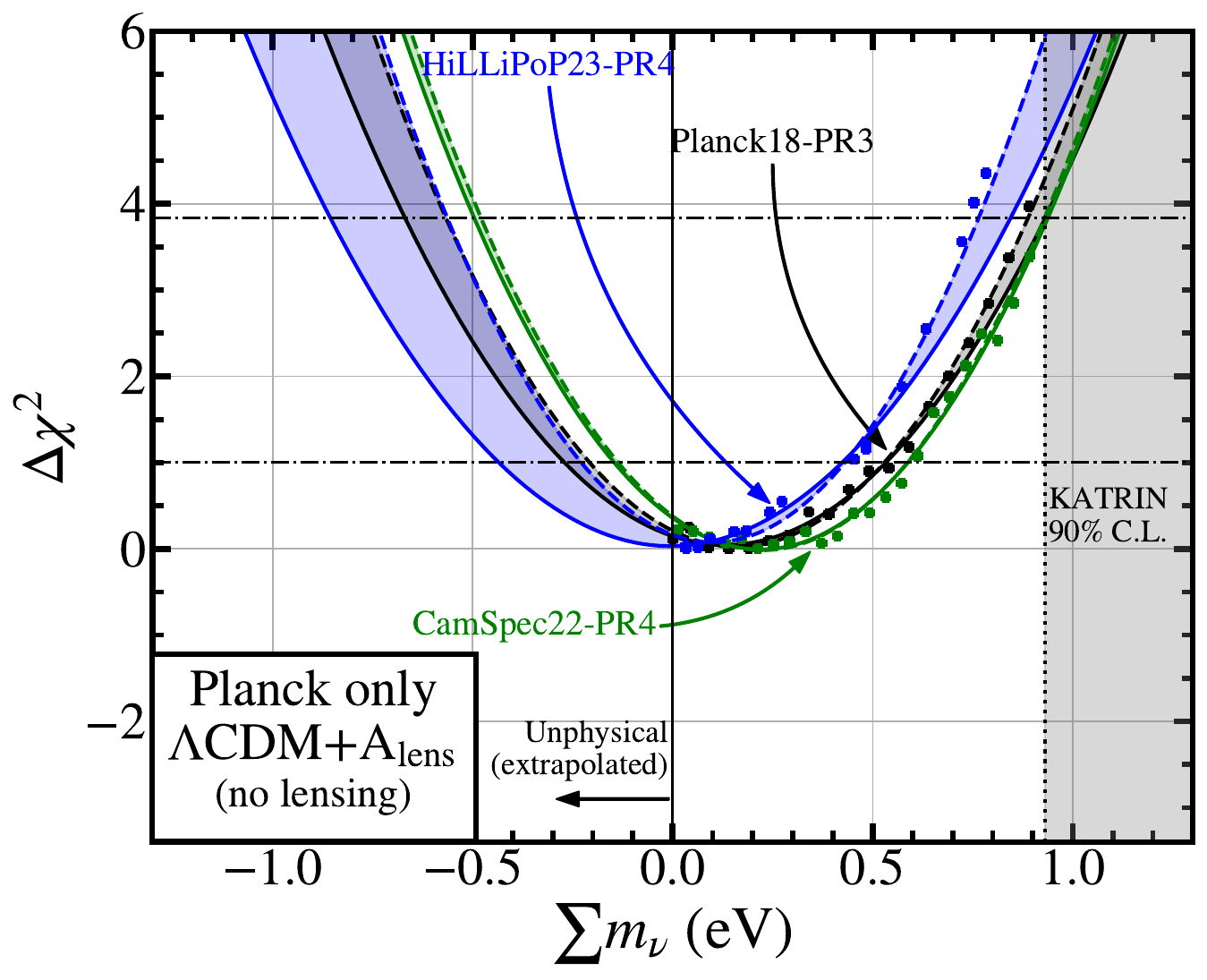}
    \caption{Neutrino mass profile likelihoods using the full Planck temperature and polarization data for $\Lambda$CDM allowing to vary the unphysical $A_{\rm lens}$ parameter which is strongly correlated with $\sum m_\nu$. We can see that the bounds are significantly relaxed and comparable to the KATRIN laboratory upper limit.}
    \label{fig:mnu_planck_Alens}
\end{figure}

\noindent \emph{The impact of $A_{\rm lens}$} --  To investigate further the potential preference for negative neutrino masses and the role of the lensing anomaly in driving this preference, we show in Fig.~\ref{fig:mnu_planck_Alens} the results from analyses that vary in addition the $A_{\rm lens}$ parameter that controls the lensing of temperature and polarization fluctuations ($C^{\phi\phi}_L = A_{\rm lens} C_{L}^{\phi\phi}|_{\Lambda \rm CDM }$ \cite{Calabrese:2008rt}). This parameter is well known to be correlated with the neutrino mass \cite{Planck:2018vyg}, and this is confirmed through our frequentist analysis, see the right panel of Fig.~\ref{fig:planck_mnu_w0_correlation}. From Fig.~\ref{fig:mnu_planck_Alens} one can see that the $\chi^2$ parabolas become all more or less similar, and that when the $A_{\rm lens}$ parameter is allowed to vary we find no preference for a negative neutrino mass. Therefore, our sampling of the $\Delta \chi^2$ in the physical region is now closer to the minimum and we can reliably extrapolate and derive robust frequentist confidence levels also for \texttt{Planck18-PR3} and \texttt{CamSpec22-PR4}. These are reported in Table~\ref{tab:Planckonly}. We see that the neutrino mass bounds can be relaxed by up to a factor of $\sim 2$ when the $A_{\rm lens}$ parameter is introduced, becoming comparable to the latest laboratory bounds. This is in agreement with the findings of~\cite{DiValentino:2021imh}. Importantly, when the $A_{\rm lens}$ parameter is allowed to vary, the profile likelihoods resemble Gaussians and we are able to obtain frequentist limits for $\sum m_\nu$. Direct comparison between Bayesian and frequentist limits shows a $10-20\%$ agreement depending upon the specific likelihood.

Though the $A_{\rm lens}$ parameter is unphysical, this exercise clearly highlights the importance of internal inconsistencies in Planck CMB data for inferences of the neutrino mass in cosmology. Marginalizing over these anomalies, it is remarkable that Planck constrains become only as strong as direct laboratory bounds. However, we stress that this is only part of the cosmological constraining power, as we have ignored the \texttt{Lensing-PR3} likelihood, which provide additional sensitivity to CMB lensing, as well as BAO and SNe data that are sensitive to the background effects of neutrinos.

\begin{figure}[t]
    \centering
    \includegraphics[width=0.498\textwidth]{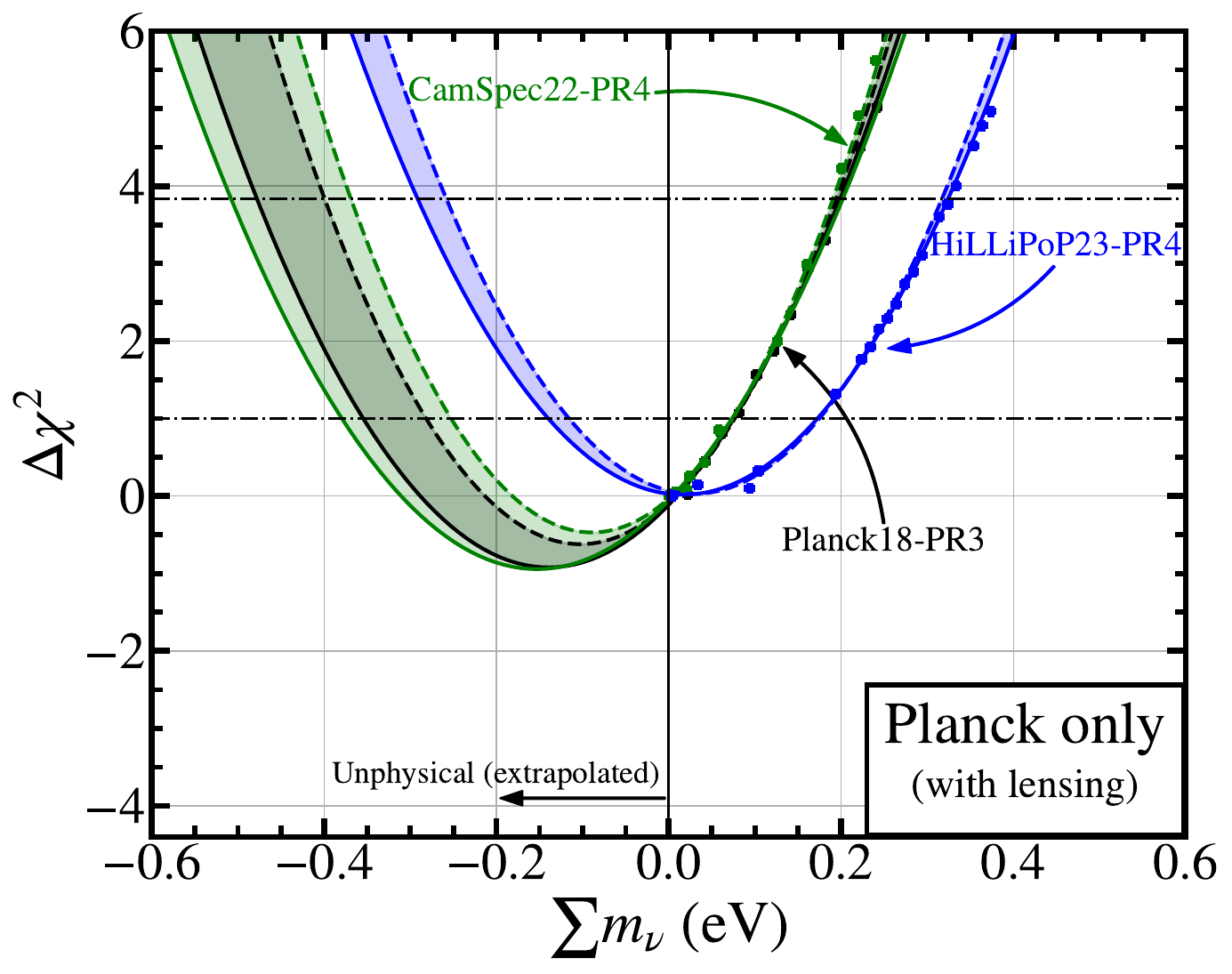}
    \caption{Neutrino mass profile likelihoods using the full Planck temperature, polarization and lensing data for $\Lambda$CDM. This should be compared with Fig.~\ref{fig:mnu_planck_only} that does not include lensing.}
    \label{fig:mnu_planck_lensing}
\end{figure}

In Fig.~\ref{fig:mnu_planck_lensing} we show the results but now including the \texttt{Lensing-PR3} lensing likelihood and with the physical condition $A_{\rm lens} = 1$. By comparing the results with those without lensing in Fig.~\ref{fig:mnu_planck_only} we can clearly see that while the impact on the \texttt{Planck18-PR3} analysis is mild at the level of the $2\sigma$ limit, the extrapolated behaviour to negative neutrino masses shows significantly weaker support for a negative best fit. This is because the lensing likelihood does not feature a lensing anomaly. The shift for the  \texttt{CamSpec22-PR4} and \texttt{HiLLiPoP23-PR4} cases is substantially more pronounced and as shown in Table~\ref{tab:Planckonly} the 95\% CL limit improves by a factor of 1.5 when the lensing likelihood is added. We note that for the data set combination \texttt{HiLLiPoP23-PR4}+\texttt{Lensing-PR3} the minimization procedure was rather challenging highlighting that a non-global minimum of the $\chi^2$ at around $\sum m_\nu \sim 0.2\,{\rm eV}$ may be present. In Appendix~\ref{app:posteriors} we show the Bayesian posterior in Fig.~\ref{fig:planck_lens_posterior} and there is a priori no evidence for multi-modality in it.

\begin{table}
\begin{center}
\begin{tabular}{|c|c|c|c|c|}
    \hline
    \multicolumn{5}{|c|}{\textbf{Planck only} $95\%$ C.L. $\sum m_\nu$ (eV)}\\
    \hline
    \multirow{2}{*}{Model}&\multirow{2}{*}{Planck likelihood}&\multirow{2}{*}{Bayesian} & \multicolumn{2}{c|}{Frequentist}\\\cline{4-5}
         &&&B.L.&F.C.\\\hline\hline
         
         \multicolumn{5}{|c|}{\textbf{No lensing likelihood}}\\\hline
         \multirow{3}{*}{$\Lambda$CDM}&\texttt{Planck18-PR3}&0.24  & 0.17&-\\\cline{2-5}
         &\texttt{\texttt{CamSpec22-PR4}}&0.33 & 0.28&-\\\cline{2-5}
         &\texttt{HiLLiPoP23-PR4}&0.51 &0.40 &0.39\\\hline
         
         \multirow{3}{*}{$\Lambda$CDM+$A_{\rm lens}$}&\texttt{Planck18-PR3}& 0.82 & 0.91&0.88\\\cline{2-5}
         &\texttt{CamSpec22-PR4} &0.78 & 0.97&0.93\\\cline{2-5}
         &\texttt{HiLLiPoP23-PR4}&0.68 &0.77 &0.76\\\hline\hline

         \multicolumn{5}{|c|}{\textbf{With lensing likelihood}}\\\hline
         \multirow{3}{*}{$\Lambda$CDM}&\texttt{Planck18-PR3}&0.25  & 0.20&0.18\\\cline{2-5}
         &\texttt{\texttt{CamSpec22-PR4}}&0.21 & 0.19&0.18\\\cline{2-5}
         &\texttt{HiLLiPoP23-PR4}&0.34 &0.33 &0.32\\\hline
    \end{tabular}
\end{center}
    \caption{
    Upper limits at 95\% CL on the neutrino mass within $\Lambda$CDM using various Planck likelihoods. We show the Bayesian limits compared with the two frequentist approaches (B.L. = Bounded Likelihood, and F.C. = Feldman-Cousins) described in Section~\ref{sec:Methods}. We do not quote frequentist F.C. bounds for \texttt{Planck18-PR3} and \texttt{CamSpec22-PR4} (no lensing) due to their non-Gaussian behaviour in the physical region, which precludes a reliable extrapolation into the unphysical regime.
    }\label{tab:Planckonly}
\end{table}

%%%%%%%%%%%%%%%%%%%%%%%%%%%%%%%%%%%%%%%%%%%%%%%%%%%%%%%%%%%%%%%%%%%%%%%%%%%%
\subsection{Planck + BAO:\\ A close look at the DESI results}\label{sec:planckbao}
%%%%%%%%%%%%%%%%%%%%%%%%%%%%%%%%%%%%%%%%%%%%%%%%%%%%%%%%%%%%%%%%%%%%%%%%%%%%
As discussed in Section~\ref{sec:nuincosmo}, there are two critical effects of the neutrino mass in the CMB: one is on CMB lensing at small angular scales (high $\ell$), see Fig.~\ref{fig:CLTTPlanck}, and the other is on the angular diameter distance to recombination, which lead to a strong correlation with other cosmological parameters such as $\Omega_m$, see Fig.~\ref{fig:mnu_Omegam_vs_H0rd}. This is why including the \texttt{Lensing-PR3} likelihood and BAO data is critical, as they can help break degeneracies between $\sum m_\nu$ and other cosmological parameters, and allow for an increased sensitivity of the neutrino mass when combined with CMB observations (note that geometric BAO data on their own are not sensitive to the neutrino mass).

\begin{figure}
    \centering
    \includegraphics[width=0.8\columnwidth]{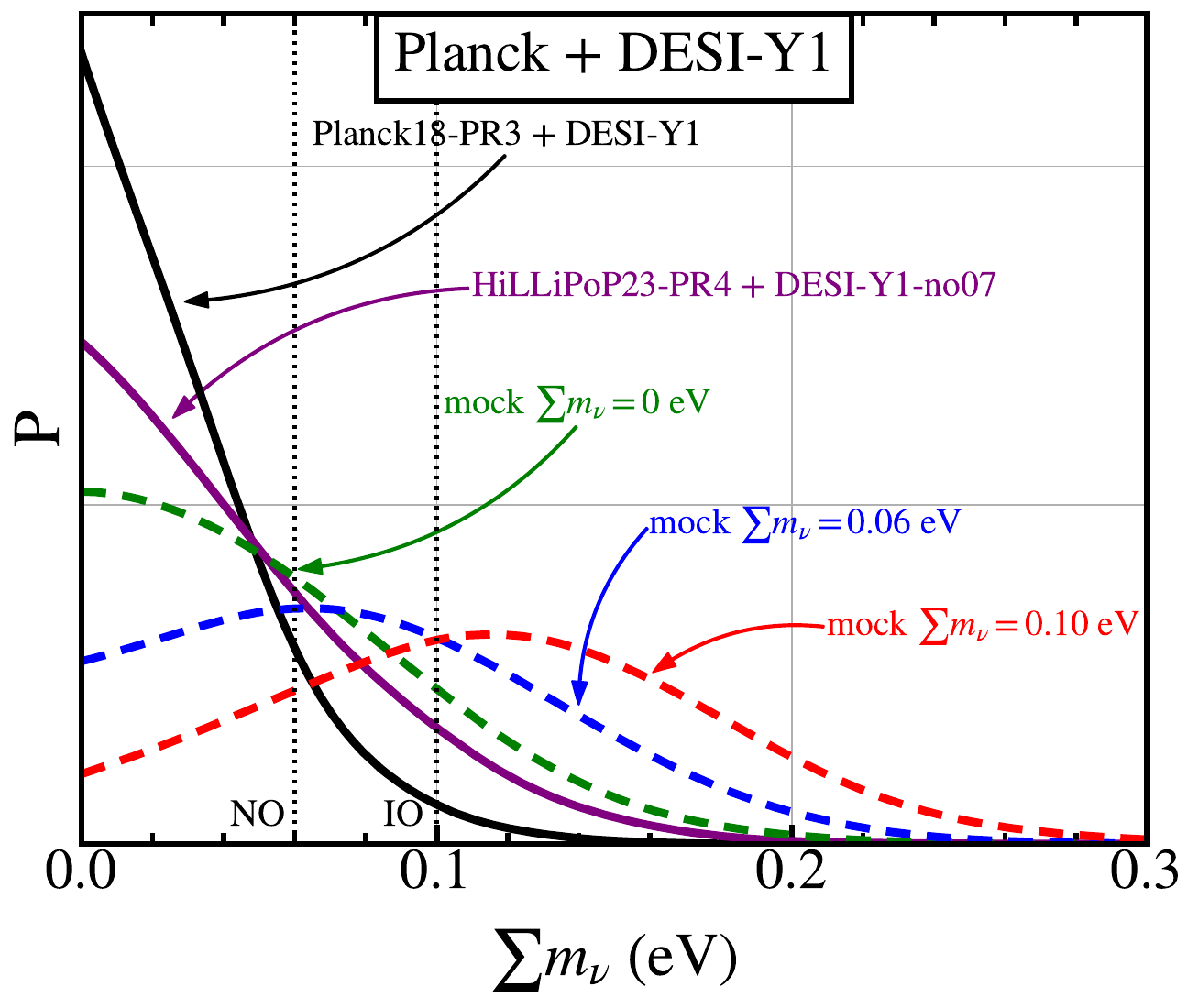}
    \caption{Reconstructed posterior distribution of $\sum m_\nu$ in a mock analysis of \texttt{Planck18-PR3}+Lensing+BAO with three configurations: $\sum m_\nu =0,0.06,0.1$ eV. We also show the posterior reconstructed from the real data for comparison.}
    \label{fig:mock_data}
\end{figure}

\begin{figure*}
    \centering
    \includegraphics[width=0.96\textwidth]{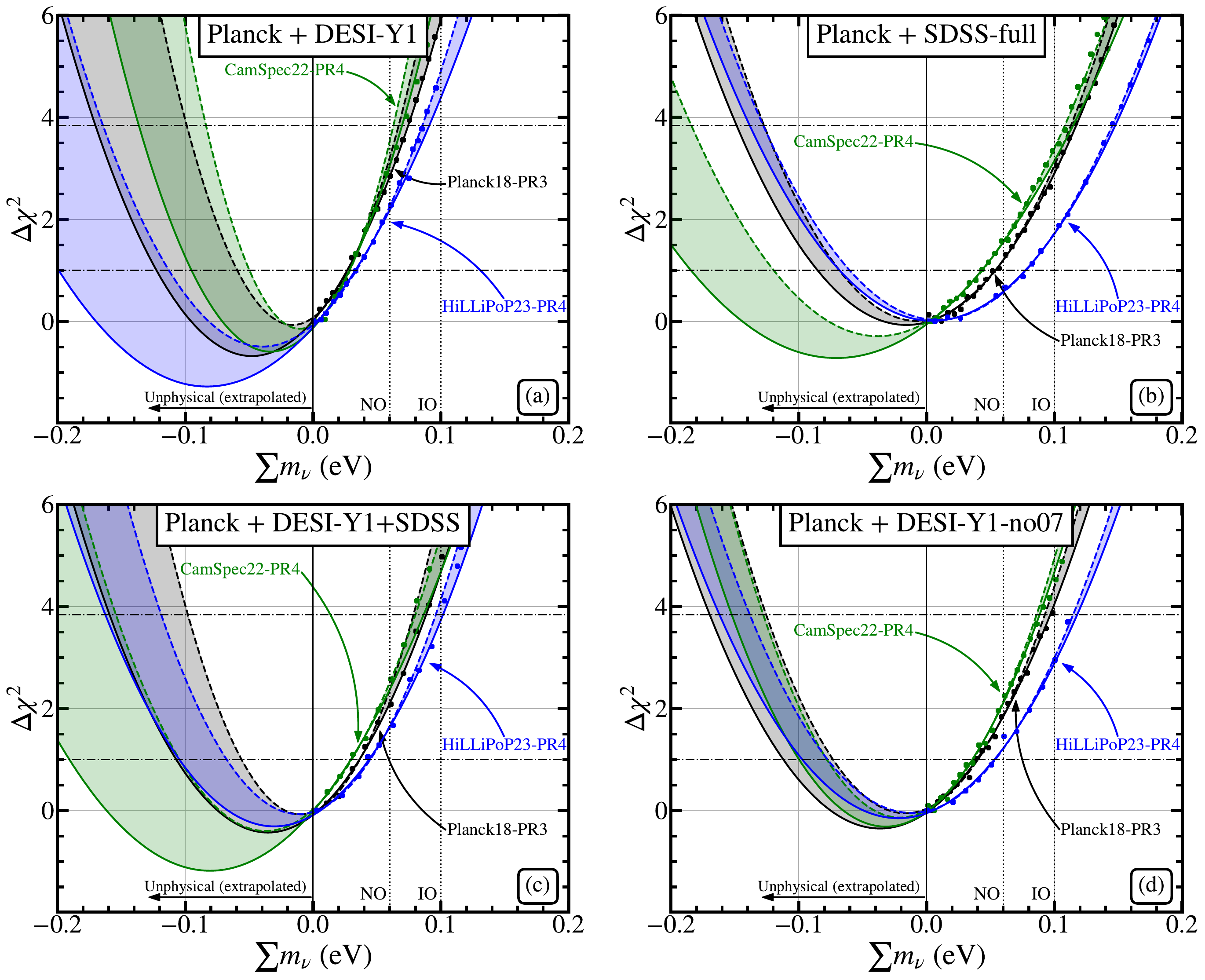}
    \caption{Profile likelihoods of the neutrino mass within $\Lambda$CDM when using Planck+BAO data comparing scenarios with DESI-Y1 (a), the full SDSS data (b), the combination of DESI-Y1+SDSS (c), and the DESI-Y1 data set but removing the ouliers at $z=0.7$ (d). We can clearly notice that systematically the HiLLiPoP23 likelihood implementation gives the weakest constraints, and that the two BAO measurements at $z = 0.7$ have a significant impact on both the bound on the neutrino mass as well as the potential preference for a negative best fit.}
    \label{fig:planck_BAO}
\end{figure*}

In particular, as stated in the introduction, one of the main goals of our study is to understand why the bound from Planck+DESI-Y1 is so strong, and whether it really points to negative neutrino masses.

In order to gauge how unexpected is the result from Planck+DESI-Y1, it is instructive to first estimate the sensitivity of this data combination to neutrino mass, as for instance done in Ref.~\cite{Brinckmann:2018owf}. To do so, we perform a mock data analysis, making use of the \texttt{MontePython} and the \texttt{Fake$\_$planck$\_$realistic} likelihood~\cite{Brinckmann:2018owf}, that we combine with our own mock likelihood of DESI-Y1, that simply makes use of the covariance matrix from the data, replacing the measurements with a fiducial model prediction. For the fiducial, we study three configurations: $\sum m_\nu =0,0.06,0.1$ eV, fixing the other $\Lambda$CDM parameters as reconstructed from \texttt{Planck18-PR3}+Lensing+BAO \cite{Planck:2018vyg}. We run MCMC chains, and show the reconstructed posteriors in Fig.~\ref{fig:mock_data}. It is clear that the real data (black line) provide significantly stronger constraints than expected, and in fact, our mock data suggests that the combination of Planck+DESI-Y1 has the sensitivity necessary to detect at small significance non-zero neutrino masses, assuming these respect the laboratory lower limits. Barring issue with our mock likelihoods, it is thus clear that there is some `anomaly' in the real data (whether due to a fluke, or new physics). As already discussed and as we will elaborate in the following with analyses of the real data, the two results that seem to be driving this unexpected constraints beyond the sensitivity are Planck likelihoods through the lensing anomaly and the DESI-Y1 data at $z=0.7$. Indeed, when we compare instead the posterior for HiLLiPoP, with a significantly reduced lensing anomaly, and remove the DESI outliers, the posterior is much closer to the expected sensitivity as shown by the purple line in Fig.~\ref{fig:mock_data}. When in presence of such anomalous behaviour from the actual data when compared to the expected sensitivities, an alternative method to the Feldman-Cousins analysis was proposed by Lokhov and Tkachov~\cite{Lokhov:2014zna} ensuring also proper coverage when deriving the confidence intervals but conservatively replacing the anomalous results pushing into the unphysical region with the expected sensitivity in absence of a signal. Since we find that the main contributors driving the anomalous results can be identified and that their impact can be almost entirely removed for instance through the HiLLiPoP + DESI-Y1 combination, we prefer not to implement this method as it would hide the different behaviours of the different datasets and cosmological models we want to analize and compare here. However, we do find the comparison with the expected sensitivities presented in Fig.~\ref{fig:mock_data}, on which the LT method is based, very illustrating. Furthermore, we can compare our expected sensitivities of Planck+DESI-Y1 to the ones obtained in previous forecasts using the full DESI reach~\cite{Brinckmann:2018owf}. We find a $1\sigma$ sensitivity of $\sim\!0.06\,{\rm eV}$ while for Planck+DESI-full Ref.~\cite{Brinckmann:2018owf} reports a $1\sigma$ sensitivity of $0.04\,{\rm eV}$. This highlights what is the level of improvement expected from the additional 5 years of BAO data, although an even higher sensitivity can of course be obtained from a full shape analysis of the matter power spectrum, see~\cite{Font-Ribera:2013rwa,EuclidTheoryWorkingGroup:2012gxx,DESI:2016fyo,Chudaykin:2019ock}.

We now turn to the real data.
%To understand this, i
In Fig.~\ref{fig:planck_BAO} we show the analyses of Planck data\footnote{We stress that \texttt{Lensing-PR3} is now included in the analysis.} combined with: (a) DESI-Y1 BAO results, (b) the full SDSS combination (which have a similar statistical power in the $\Omega_m$ vs $H_0r_d$ plane as DESI-Y1, see Fig.~\ref{fig:mnu_Omegam_vs_H0rd}), (c) the DESI/SDSS BAO combination that takes the BAO results at each redshift bin from the survey that has greatest statistical power, and (d) the DESI-Y1 BAO combination but without the $z = 0.7$ bin that contains data which deviate by $\sim 3\sigma$ from $\Lambda$CDM predictions. Importantly, in all data set combinations, we consider separately the three latest Planck likelihood implementations.

First, from panel (a), we confirm that the extrapolation of  the results from Planck+DESI-Y1 to the unphysical region seem to favor negative masses. However, in our approach, given uncertainties in the profile represented by the bands, the preference remains at low statistical significance. In fact, we find that it is the combination with \texttt{HiLLiPoP23-PR4} that seems to lead to the largest preference (reaching at most $\sim 1\sigma$). This may appear counter-intuitive given results presented in previous section. We attribute this to the fact that \texttt{HiLLiPoP23-PR4} has the weakest constraining power on neutrino masses, and thus the statistical power from DESI can be more pronounced. 
Second, by comparing the results from Planck+DESI-Y1 and Planck+SDSS in panel (b), one can clearly see that the bound on the neutrino mass are significantly weaker in the latter case and that the preference for negative neutrino masses further decrease. This is particularly striking for \texttt{HiLLiPoP23-PR4} that is now fully compatible with positive masses.
%is significantly stronger for the DESI-Y1 combination and that it is precisely in this scenario that extrapolation to the unphysical regime leads to a negative best-fit. 
However, when comparing the Planck+DESI-Y1 and Planck+DESI-Y1+SDSS (panel (c)) one notices that the profiles are quite similar and only slightly shifted towards more positive values for the latter case. This suggests that the low-$z$ data points from DESI, that are replaced by SDSS is this analysis, do not play a significant role in the preference for negative neutrino masses. Rather, we can compare the results from Planck+DESI-Y1 and Planck+DESI-Y1no07 (panel (d)) which excludes the outliers at $z=0.7$. We find that removing these data points loosens the bound and also makes all likelihood combinations to peak at around $\sum m_\nu \simeq 0$, thus removing the potential preference for negative neutrino masses. Although it is obvious that removing two data points from DESI-Y1 is not a valid statistical procedure, this exercise shows that DESI-Y1 BAO data at $z= 0.7$ have a very significant impact on the neutrino mass bound. We note that the DESI-Y1+SDSS BAO combination does include these outliers in the combination as the effective volume of DESI is already larger than that of SDSS at $z = 0.7$.

Finally, although Fig.~\ref{fig:planck_BAO} shows that the form of the profiles (in particular the extrapolation) is different for the three Planck implementations, the differences at the level of the 95\% CL bound is only at the $10-20\%$ level, depending upon the specific BAO data used. 
This is good news because it shows that the potentially large systematic shifts on $\sum m_\nu$ when using Planck CMB data alone are not present at the same level when CMB lensing and geometric BAO data are included. Our results are in qualitative agreement with the recent Bayesian analysis of~\cite{Allali:2024aiv}.

%%%%%%%%%%%%%%%%%%%%%%%%%%%%%%%%%%%%%%%%%%%%%%%%%%%%%%%%%%%%%%%%%%%%%%%%%%%%
\subsection{Planck + DESI + SN: Impact of the Dark Energy equation of state}\label{sec:planckbaoSN}
%%%%%%%%%%%%%%%%%%%%%%%%%%%%%%%%%%%%%%%%%%%%%%%%%%%%%%%%%%%%%%%%%%%%%%%%%%%%

\begin{figure*}[t]
    \centering
    \includegraphics[width=1\textwidth]{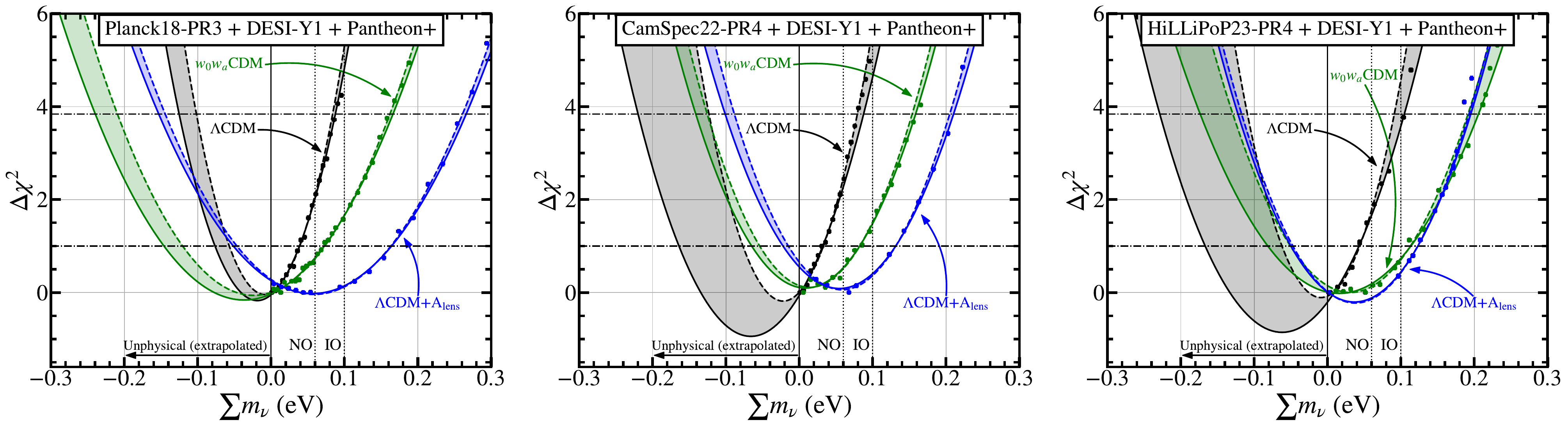}
    \caption{Neutrino mass profile likelihoods for Planck+DESI-Y1+Pantheon+ data set combinations. We show $\Lambda$CDM in black, varying the equation of state of dark energy in green, and allowing for $A_{\rm lens}$ to vary in blue. In the left panel we show the results for plik, in the middle for CamSpec, and in the right panel for Hillipop. We clearly see a similar behaviour for all of them and the potential preference for a negative best fit to dissapear when the equation of state of dark energy is allowed to vary.}
    \label{fig:planck_MAX}
\end{figure*}

 So far, we have explored constraints to $\sum m_\nu$ assuming a flat $\Lambda$CDM background, and showed how BAO data can help strengthen the bound by breaking the degeneracy with $\Omega_m$. However, in models with more parameters controlling the late-time expansion history, it is expected that additional degeneracies with the neutrino masses will appear. Chief amongst those is the well-known degeneracy with the equation of state of dark energy, $w$ (see e.g.~\cite{Boyle:2017lzt}.) 
 Given the tentative evidence for a time-varying equation of state of dark energy from DESI when combined with SN data \cite{DESI:2024mwx}, it is relevant to explore how it can impact the bound on the neutrino masses. 
 Following \cite{DESI:2024mwx}, we model the equation of state of dark energy to vary according to the Chevalier-Polarski-Linder parametrization $w(a) = w_0 + (1-a)w_a$~\cite{Chevallier:2000qy,Linder:2002et}, where $a$ is the scale factor, and vary $w_0 \in [-3,2]$ and $w_a\in [-3,1]$.

In Fig.~\ref{fig:planck_MAX} we show the likelihood profile of $\sum m_\nu$ built from the combination of Planck, DESI-Y1 BAO and the SN Pantheon sample, in $\Lambda$CDM (in black) and in the $w_0w_a$CDM cosmology (in green). One can see that when the equation of state of dark energy is allowed to vary the bound on the neutrino mass is relaxed, in good agreement with \cite{DESI:2024mwx}. We find, however, that the 95\% bound can be roughly 30\% weaker for HiLLiPoP than with Plik or CamSpec. Interestingly we also note that the potential evidence for a negative neutrino mass vanishes. 

Something important to highlight is that, while allowing for the  equation of state of dark energy to vary relaxes the bound on the neutrino mass, the best fit for the equation of state differs significantly from the cosmological constant value and in particular suggests that $w_0 \gtrsim -1$ today. Thus, while the bound may be relaxed, large neutrino masses would require dark energy to behave very differently than a cosmological constant. We note that if one restricts the analysis to constant equation of state, namely $w(a)=w_0$ where only $w_0$ is allowed to vary, the bound on the neutrino mass remains very similar to that in $\Lambda$CDM, see~\cite{DESI:2024mwx}.

Lastly, we investigate whether varying the $A_{\rm lens}$ parameter may remove the preference for negative neutrino masses, despite the inclusion of DESI-Y1 BAO and SN data data. Our results are shown in Fig.~\ref{fig:planck_MAX} in blue. One can notice that this shifts the best fit to the positive regime and that the bound becomes again a factor of $\sim 2$ weaker than when compared to the standard case where $A_{\rm lens} = 1$.
This suggests that, regardless of the behavior of DESI data, it is the lensing anomaly that dominates the preference for negative neutrino masses. Note though, that removing (most of) the constraining power from lensing by including $A_{\rm lens}$ does not remove all the sensitivity to neutrino masses altogether. While the constraints relax, they remain significantly stronger than laboratory ones when BAO and SNIa data are included, in the ball park of $\sum m_\nu \lesssim 0.2-0.3$ eV depending on which CMB likelihood is used.

%%%%%%%%%%%%%%%%%%%%%%%%%%%%%%%%%%%%%%%%%%%%%%%%%%%%%%%%%%%%%%%%%%%%%%%%%%%%
\subsection{Frequentists vs Bayesian Limits: The impact of statistical choices on the neutrino mass bound}\label{sec:statistics}
%%%%%%%%%%%%%%%%%%%%%%%%%%%%%%%%%%%%%%%%%%%%%%%%%%%%%%%%%%%%%%%%%%%%%%%%%%%%

Until now, we have focused our attention on profile likelihoods, as these allowed us to investigate the preference for negative neutrino masses. In this section we will study how the confidence intervals built from the profile likelihood compare with the Bayesian credible intervals built from the posteriors. While the two statistical approaches address distinct questions and thus do not need to necessarily coincide, we find the comparison useful to understand how dependent is the derived constraint on the choice of the statistical procedure. Moreover, if the likelihood is Gaussian and when assuming flat, non-informative priors on the parameter under study, the two approaches should coincide. Therefore, the comparison allows to asses the role of possible prior effects in the bound given by Eq.~\eqref{eq:boundDESI} as well as deviations from Gaussianity. Notice that the extrapolations to negative masses performed in our work as well as in Ref.~\cite{eBOSS:2020yzd} and~\cite{Allali:2024aiv} rely on the Gaussianity of the likelihood and posterior respectively. Similarly, the study of Ref.~\cite{Elbers:2024sha} relies on a linear expansion of the dependence of the observables with $\sum m_\nu$ which, per Wilks theorem, would imply a Gaussian behaviour. The fact that these analyses do not coincide (see section~\ref{sec:negativeneutrinos} for a full comparison) implies that the Gaussian behaviour approximation is violated at some degree. As we will show and discuss below, we find that the differences between the Bayesian and frequentist approaches are at the $10\%$ level, which provides a useful estimation of the size of the uncertainties introduced by these approximations. We refer to Appendix~\ref{app:posteriors} for the posterior distributions. 

%It remains to be seen however, how the confidence intervals built from the profile likelihood compares with the Bayesian credible intervals built from the posteriors, in order to understand how important is the choice of statistical procedure and to quantify the role of prior effect in the bound given by Eq.~\eqref{eq:boundDESI}.  

\begin{table*}
\begin{center}
    \begin{tabular}{|c|c|c|c|c|c|}
    \hline
    \multicolumn{6}{|c|}{\textbf{Planck + BAO} $95\%$ C.L. $\sum m_\nu$ (eV)}\\
    \hline
         \multirow{2}{*}{BAO data}&\multirow{2}{*}{Planck likelihood}&\multirow{2}{*}{Bayesian} & \multicolumn{2}{c|}{$\,$ Frequentist $\,$}&\multirow{2}{*}{$\,$Gaussian fit $\,$}\\\cline{4-5}
         &&&$\,$ B.L. $\,$&F.C.&\\\hline\hline
         \multirow{3}{*}{SDSS-full}&\texttt{Planck18-PR3}& 0.114 & 0.116&0.113&-0.018$\pm$0.068\\\cline{2-6}
         &\texttt{CamSpec22-PR4}& 0.115& 0.108&0.106&-0.026$\pm$0.067\\\cline{2-6}
         &\texttt{HiLLiPoP23-PR4}&0.151 & 0.146&0.146&$\,\,$0.007$\pm$0.071\\\hline
         \hline
         \multirow{3}{*}{DESI-Y1}&\texttt{Planck18-PR3}& 0.084 & 0.074&0.071&-0.047$\pm$0.057\\\cline{2-6}
         &\texttt{CamSpec22-PR4}& 0.079& 0.069&0.067&-0.045$\pm$0.053\\\cline{2-6}
         &\texttt{HiLLiPoP23-PR4}& 0.102& 0.085&0.083&-0.038$\pm$0.060\\\hline
         \hline
        \multirow{3}{*}{DESI-Y1+SDSS}&\texttt{Planck18-PR3}& 0.096&0.086 & 0.082&-0.033$\pm$0.058\\\cline{2-6}
         &\texttt{CamSpec22-PR4}&0.088& 0.080& 0.077&-0.032$\pm$0.054\\\cline{2-6}
         &\texttt{HiLLiPoP23-PR4}& 0.112&0.099&0.097 &-0.016$\pm$0.058\\\hline  
         \hline
         \multirow{3}{*}{DESI-Y1-no07}&\texttt{Planck18-PR3} &0.107&0.096&0.092&-0.036$\pm$0.065\\\cline{2-6}
         &\texttt{CamSpec22-PR4} &0.101&0.089&0.087&-0.048$\pm$0.066\\\cline{2-6}
         &\texttt{HiLLiPoP23-PR4} &0.125&0.114&0.114&-0.012$\pm$0.064\\\hline
    \end{tabular}
\end{center}
    \caption{
    Upper limits at 95\% CL on the neutrino mass within $\Lambda$CDM using various data set combinations of Planck + BAO data. We show the Bayesian limits compared with the two frequentist approaches (B.L. = Bounded Likelihood, and F.C. = Feldman-Cousins) described in Section~\ref{sec:Methods}. We also report the Gaussian fit for our profile likelihoods.
    %The Bayesian results  from~\cite{Allali:2024aiv} are: 2018 + DESI is 0.077, the one from H2020 + DESI is 0.086, the one from C2020+DESI is 0.080. Others include H2020+SDSS/DESI 0.11, H2020 + SDSSfull = 0.14.
    }\label{tab:PlanckBAO}
\end{table*}

Our results are summarized in Table~\ref{tab:PlanckBAO} for analyses that combine Planck+BAO data, and in Table~\ref{tab:PlanckBAOSN} for those that also include SN data from the Pantheon sample. These tables include three estimates of the bound to neutrino masses:  the Bayesian limit at 95\% CL, those derived using Feldman-Cousins procedure (F.C.), as well as those using the naive bounded maximum likelihood (B.L.) ($\Delta \chi^2 = 3.84$), all at the same confidence level. Firstly, we generally notice a very good agreement between the two frequentists approaches, with differences between them at the $\lesssim 5\%$ level only. This suggests that the fact that the minimum lie beyond the physical region does not significantly affect the bounds to neutrino masses. Secondly, and interestingly, we also notice a very good agreement between the frequentists and Bayesian limits. In fact, we find, that the frequentists limits are in many cases $\sim 10\%$ stronger than the Bayesian ones. For example, considering the data set combination of Plik+DESI, we find at 95\% CL:
\begin{subequations}
\begin{align}
    \sum m_\nu &< 0.084 \,{\rm eV} \,\,\,[{\rm Bayesian}]\,,\label{eq:mnuBoundPlanckDESIBayesian}\\
    \sum m_\nu &< 0.074 \,{\rm eV} \,\,\,[{\rm Bounded\!-\!Likelihood}]\,,\\
    \sum m_\nu &< 0.071 \,{\rm eV} \,\,\,[{\rm Feldman\!-\!Cousins}]\,.
\end{align}
\end{subequations}
One  clearly sees that the three are very similar, with the frequentist ones being slightly more stringent. While the two approaches need not necessarily agree, this could be due to two effects.  First, it is possible that there are mild prior effects in the Bayesian analysis, that go in the direction of relaxing the bound. Second, it can be difficult to find the absolute minimum of the $\chi^2$ for each simulated value of $\sum m_\nu$ for such a large parameter space. If the simulated annealing methods fails to cool to the absolute minimum, the slightly larger values of the $\chi^2$ would lead to slightly tighter frequentist constraints. Nevertheless and regardless of its origin, this effect is only around the $10\%$ level, and we thus conclude that the constraints are robust to the choice of statistical method up to that level of difference. Let us additionally note that the Bayesian constraint we derive here is slightly different than Eq.~\eqref{eq:boundDESI}. This is because the DESI collaboration used more constraining CMB lensing data, combining Planck lensing PR4 with ACT lensing, rather than Planck lensing PR3 as we do here.
Nevertheless, we do not expect that using this lensing data would change the overall trend.

\begin{table*}
\begin{center}
    \begin{tabular}{|c|c|c|c|c|c|}
    \hline
    \multicolumn{6}{|c|}{\textbf{Planck + DESI-Y1 + Pantheon+} $95\%$ C.L. $\sum m_\nu$ (eV)}\\
    \hline
         \multirow{2}{*}{Model}&\multirow{2}{*}{Planck likelihood}&\multirow{2}{*}{Bayesian} & \multicolumn{2}{c|}{$\,\,$Frequentist$\,\,$}&\multirow{2}{*}{$\,$Gaussian fit$\,$}\\\cline{4-5}
         &&&$\,\,$B.L.$\,\,$&F.C.&\\\hline\hline
         \multirow{3}{*}{$\Lambda$CDM}&\texttt{Planck18-PR3}& 0.093 & 0.087&0.088&-0.025$\pm$0.056\\\cline{2-6}
         &\texttt{CamSpec22-PR4} &0.089 & 0.080&0.078&-0.033$\pm$0.055\\\cline{2-6}
         &\texttt{HiLLiPoP23-PR4}&0.112 &0.099 &0.098&-0.034$\pm$0.066\\\hline
         \hline
         \multirow{3}{*}{$w_0w_a$CDM}&\texttt{Planck18-PR3}&0.177  & 0.163&0.163&-0.019$\pm$0.092\\\cline{2-6}
         &\texttt{CamSpec22-PR4}&0.167 & 0.161&0.164&$\,\,$0.006$\pm$0.079\\\cline{2-6}
         &\texttt{HiLLiPoP23-PR4}&0.213 & 0.205&0.207&$\,\,$0.024$\pm$0.092\\\hline
         \hline
         \multirow{3}{*}{$\Lambda$CDM+$A_{\rm lens}$}&\texttt{Planck18-PR3}&0.242  & 0.268&0.260&$\,\,$0.060$\pm$0.102\\\cline{2-6}
         &\texttt{\texttt{CamSpec22-PR4}}&0.204 & 0.220&0.210&$\,\,$0.050$\pm$0.079\\\cline{2-6}
         &\texttt{HiLLiPoP23-PR4}&0.180 &0.187 &0.181&$\,\,$0.046$\pm$0.068\\\hline
    \end{tabular}
    \caption{Upper limits at 95\% CL on the neutrino mass using Planck + BAO + SN data within extended $\Lambda$CDM models, including a time-varying equation of state of dark energy, as well as varying the $A_{\rm lens}$ parameter. We show the Bayesian limits compared with the two frequentist approaches (B.L. = Bounded Likelihood, and F.C. = Feldman-Cousins) described in Section~\ref{sec:Methods}. We also report the Gaussian fit for our profile likelihoods, obtained from the points below $\Delta\chi^2=4$. 
    %The Bayesian 2018 one can be compared with 0.093 at 2sigma as found in~\cite{Wang:2024hen} (CDS). The one for 2018 as reported in ~\cite{Allali:2024aiv} is 0.086, the one for H2020 is 0.099
    }\label{tab:PlanckBAOSN}
\end{center}
\end{table*}

So far we have included in our analyses values for neutrino masses down to the massless limit, $\sum m_\nu = 0$, but we know from the laboratory that there are  physical boundaries at either $\sum m_\nu = 0.06\,{\rm eV}$ for NO or at $\sum m_\nu = 0.10\,{\rm eV}$ for IO. 
To gauge the impact of those experimental lower limits on the cosmological neutrino mass bound, we run dedicated Bayesian analyses restricting the prior to $\sum m_\nu$ following either the NO or IO constraints. For the frequentist limit, it is sufficient to consider these boundaries as lower limits in our $\Delta \chi^2$ curves. This procedure yields:
\begin{subequations}
\begin{align}
    \sum m_\nu &< 0.121 \,{\rm eV} \,\,\,[{\rm NO\!-\!Bayesian}]\,,\\
    \sum m_\nu &< 0.106 \,{\rm eV} \,\,\,[{\rm NO\!-\!Bounded\!-\!Likelihood}]\,,\\
    \sum m_\nu &< 0.096 \,{\rm eV} \,\,\,[{\rm NO\!-\!Feldman\!-\!Cousins}]\,,
\end{align}
\end{subequations}
and for the inverted ordering case:
\begin{subequations}
\begin{align}
    \sum m_\nu &< 0.152 \,{\rm eV} \,\,\,[{\rm IO\!-\!Bayesian}]\,,\\
    \sum m_\nu &< 0.138 \,{\rm eV} \,\,\,[{\rm IO\!-\!Bounded\!-\!Likelihood}]\,,\\
    \sum m_\nu &< 0.127 \,{\rm eV} \,\,\,[{\rm IO\!-\!Feldman\!-\!Cousins}]\,.
\end{align}
\end{subequations}
Since in these scenarios the physical boundary is further away from the best fit of the $\Delta \chi^2$, the Feldman-Cousins correction becomes more relevant and we observe a larger difference compared to the naive bound one would derive simply assuming the applicability of Wilk's theorem, although it is still within $10\%$. The difference between the frequentist and Bayesian constraints also increases, with up to $20\%$ difference between the Feldman-Cousins result and the Bayesian posterior. Let us stress that, for this particular dataset, the inverted ordering assumption has a $p-$value of only $1\%$.

Importantly, we have highlighted before that there are two effects that significantly pull the bound on the neutrino mass in Eq.~\eqref{eq:boundDESI}: i) the lensing anomaly present in some of the Planck likelihoods, and ii) the outliers in DESI-Y1 at $z=0.7$. In this context, to be maximally conservative, one can consider the combination of HiLLiPoP+DESIY1no07 for which there is no lensing anomaly in the Planck likelihood and where the outliers in DESI-Y1 data have been removed. The relevant Bayesian and frequentist limits from HiLLiPoP+DESIY1no07 read:
\begin{subequations}
\begin{align}
    \sum m_\nu &< 0.125 \,{\rm eV} \,\,\,[{\rm Bayesian}]\,,\label{eq:mnuBoundPlanckDESIno07Bayesian}\\
    \sum m_\nu &< 0.114 \,{\rm eV} \,\,\,[{\rm Bounded\!-\!Likelihood}]\,,\\
    \sum m_\nu &< 0.114 \,{\rm eV} \,\,\,[{\rm Feldman\!-\!Cousins}]\,.
\end{align}
\end{subequations}
Here we can see again a $\sim\! 10\%$ agreement between Bayesian and frequentist approaches.

When considering the physical boundary for normal ordering, the constraints read:
\begin{subequations}
\begin{align}
    \sum m_\nu &< 0.160 \,{\rm eV} \,\,\,[{\rm NO-Bayesian}]\,,\label{eq:mnuBoundPlanckDESIno07NOBayesian}\\
    \sum m_\nu &< 0.132 \,{\rm eV} \,\,\,[{\rm NO-Bounded\!-\!Likelihood}]\,,\\
    \sum m_\nu &< 0.125 \,{\rm eV} \,\,\,[{\rm NO-Feldman\!-\!Cousins}]\,.
\end{align}
\end{subequations}
and for the inverted ordering case:
\begin{subequations}
\begin{align}
    \sum m_\nu &< 0.179 \,{\rm eV} \,\,\,[{\rm IO-Bayesian}]\,,\label{eq:mnuBoundPlanckDESIno07IOBayesian}\\
    \sum m_\nu &< 0.156 \,{\rm eV} \,\,\,[{\rm IO-Bounded\!-\!Likelihood}]\,,\\
    \sum m_\nu &< 0.146 \,{\rm eV} \,\,\,[{\rm IO-Feldman\!-\!Cousins}]\,.
\end{align}
\end{subequations}
Where, as before, the FC corrections become more relevant given the larger distance between the best fit and the physical boundary.

Finally, we can compare our Bayesian bounds with other recent studies. In particular, our limit for the PlanckPR3+DESI+Pantheon perfectly agrees with the one reported in~\cite{Wang:2024hen}. Ref.~\cite{Allali:2024aiv} also presented analyses including various versions of the new Planck likelihoods. For the data combination Plik+DESI and HiLLiPoP+DESI (with or without SN), we find bounds that are $\sim 10-20\%$ looser that those reported in~\cite{Allali:2024aiv}. However, for the case HiLLiPoP+SDSS/DESI we find the same limit as~\cite{Allali:2024aiv}. Given that we agree with Ref.~\cite{Wang:2024hen} when the very same data is considered, but also with Ref.~\cite{Allali:2024aiv} when a subset of the DESI data set is considered, we conjecture that the differences in the limits may stem from a different implementation of the full DESI likelihood. Our implementation matches the one in the Cobaya public repository~\cite{Torrado:2020dgo}.

%%%%%%%%%%%%%%%%%%%%%%%%%%%%%%%%%%%%%%%%%%%%%%%%%%%%%%%%%%%%%%%%%%%%%%%%%%%%%%%%%%%%%%%%%%%
\section{Do cosmological data prefer a ``negative" neutrino mass?}\label{sec:negativeneutrinos}
%%%%%%%%%%%%%%%%%%%%%%%%%%%%%%%%%%%%%%%%%%%%%%%%%%%%%%%%%%%%%%%%%%%%%%%%%%%%%%%%%%%%%%%%%%%

\begin{figure*}[t]
    \centering
    \includegraphics[width= 0.8\textwidth]{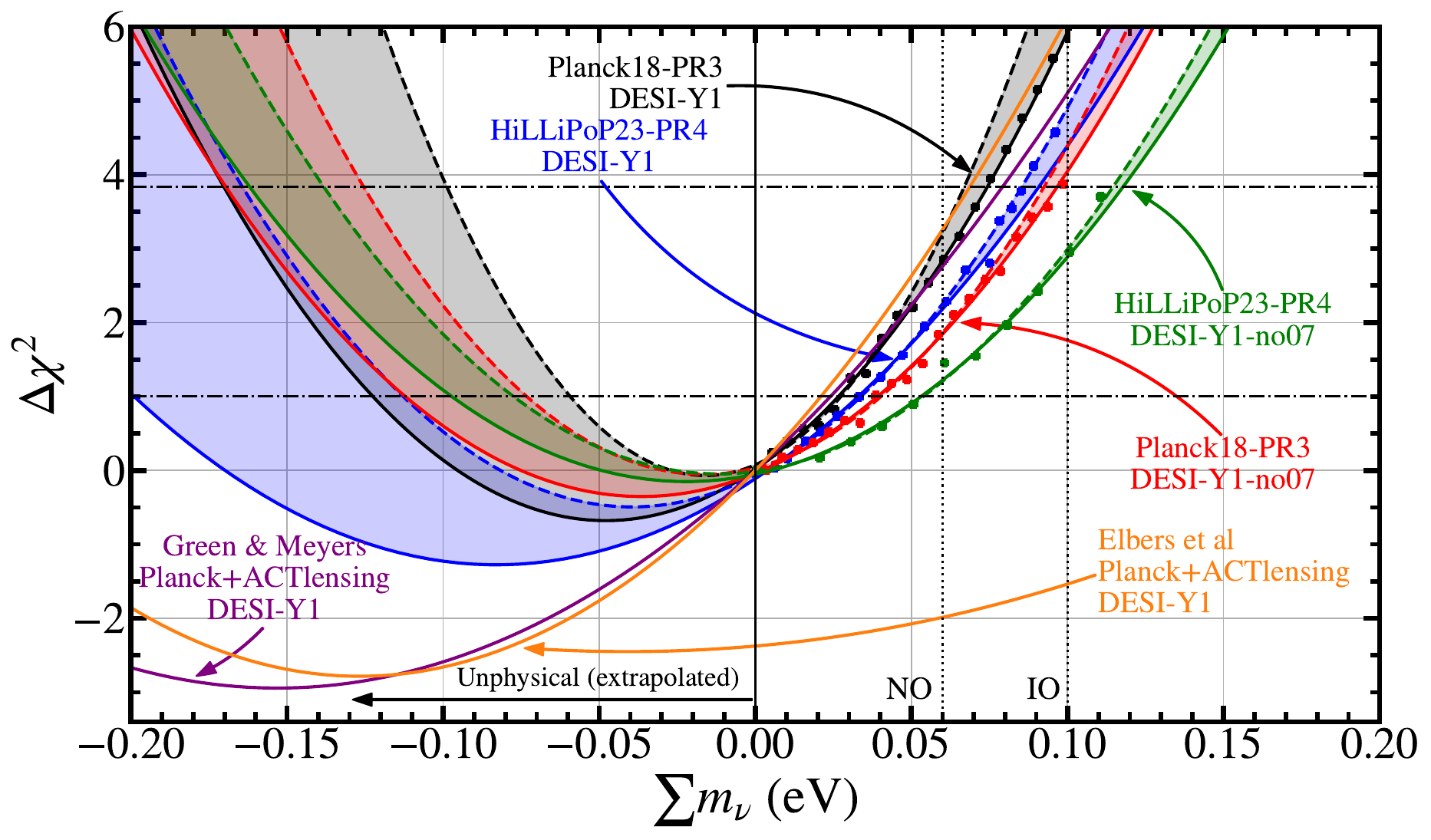}
    \caption{Profile likelihoods for the data set combinations of Planck18-PR3+DESI-Y1 (black), HiLLiPoP23-PR4 (blue), Planck18-PR3+DESI-Y1-no07 (red), HiLLiPoP23-PR4+DESI-Y1-no07 (green), and compared with $\chi^2_{\rm eff}=-2\log \mathcal{P}$ from~\cite{Green:2024xbb} (purple) and~\cite{Elbers:2024sha} (orange), which have treatments for ``negative" neutrino masses. By comparing the black and blue curves we can clearly see that the bound on the neutrino mass gets relaxed if the HiLLiPoP likelihood (which does not contain a lensing anomaly) is used. However, it is clear from the extrapolated parabolas that there is still some preference for a negative neutrino mass. This, however, disappears when the DESI BAO data at $z =0.7$ which contains a $\sim 3\sigma$ outlier is removed (see red and green curves).}
    \label{fig:mnuPlanckBAO_comparison}
\end{figure*}

It has been recently emphasized in~\cite{Craig:2024tky,Green:2024xbb,Elbers:2024sha} that cosmological data may prefer ``negative" neutrino masses and that such a preference for unphysical values, if not due to a statistical fluke or an unknown systematic effect, may hint for new physics in cosmology. 
In their analysis, the Authors of~\cite{Craig:2024tky,Green:2024xbb} define an ``effective neutrino mass'' which, when taking negative values, generates an {\it increase} of power in the CMB lensing potential, an effect that is opposite to that of positive neutrino masses. In~\cite{Elbers:2024sha} an alternative definition is adopted which also accounts for the other effects of neutrino masses in cosmological observables (in particular on the angular diameter distance, but not only). In this paper, to gauge the preference for neutrino mass, we have simply extrapolated through a Gaussian fit the behaviour of our profiled $\Delta \chi^2$. Even though the extrapolation introduces some degree of uncertainty, that we have quantified with the procedure detailed in Section \ref{sec:Methods}, {it provides a complementary way to assess the preference for negative values displayed by the different datasets analyzed without the need for an explicit (arbitrary) modelling of the effect of negative neutrino masses. A similar approach, extrapolating instead a Gaussian fit to the posterior distribution, was adopted in Refs.~\cite{eBOSS:2020yzd} and~\cite{Allali:2024aiv}.

Our results agree overall fairly well with all previous works~\cite{Craig:2024tky,Allali:2024aiv,Green:2024xbb,Elbers:2024sha} considering the different treatments performed in each of them. Our main results and the comparison to previous analyses are shown in Fig.~\ref{fig:mnuPlanckBAO_comparison} (see also App.~\ref{app:comparison}). In particular, Ref.~\cite{Elbers:2024sha} argued that the Gaussian extrapolation in~\cite{Allali:2024aiv} underestimates the preference for negative masses of present data. Indeed, Refs.~\cite{Craig:2024tky,Green:2024xbb,Elbers:2024sha} find rather more negative best fits through their analyses with their respective ``effective masses'' peaking around $ -0.15\,{\rm eV}$ (see purple and orange lines in Fig.~\ref{fig:mnuPlanckBAO_comparison}). We do find some preference for negative neutrino masses in datasets including both the Planck likelihoods affected by the lensing anomaly (Plik 2018 in Fig.~\ref{fig:mnuPlanckBAO_comparison} but also CamSpec in previous sections) and the full DESI Y1 BAO data (black line). However, this preference is significantly weaker, with a best fit at $\sum m_\nu \simeq -0.05\,{\rm eV}$ in agreement with the extrapolation of the posterior shown in~\cite{Allali:2024aiv} and the comparison and discussion in~\cite{Elbers:2024sha}.

More importantly, Ref.~\cite{Elbers:2024sha} only considered the CamSpec Planck likelihood. We find here that this dataset has a preference for negative neutrino masses, because of the presence of a residual lensing anomaly, and that it can be reduced when using alternative likelihoods. Indeed, we find a preference for negative neutrino masses when Planck 2018 is analyzed on its own. Switching for the \texttt{HiLLiPoP23-PR4} version of Planck (largely unaffected by the lensing anomaly), there is no preference for negative masses, and the bound significantly relaxes. The preference re-appears for \texttt{HiLLiPoP23-PR4} when DESI-Y1 BAO data are included because these data have a lower constraining power to $\sum m_\nu$. Nevertheless, Fig.~\ref{fig:mnuPlanckBAO_comparison} shows that the bound on $\sum m_\nu$ relaxes when changing from plik 2018 to HiLLiPoP (black vs blue lines) or when removing the $z=0.7$ outlier bin in DESI (black vs red lines). Finally, when the $z=0.7$ bin is removed and the HiLLiPoP implementation are both considered, the preference for negative masses essentially disappears (green line) and the bound is about a factor 2 weaker. Let us note that the authors of~\cite{Green:2024xbb} did investigate the persistence of their effective negative neutrino mass signal when PR4 likelihoods are used. However, the analyses were not performed for \texttt{HiLLiPoP23-PR4} but rather for the \texttt{CamSpec22-PR4} likelihood, where the lensing anomaly is still present, albeit at a reduced level. Ref.~\cite{Green:2024xbb} does show that the preference for negative masses is relaxed in this scenario at a level that, naively, seems consistent with the reduced presence of the anomaly in CamSpec. 
In this context, it would be very interesting to explore the potential preference for a ``negative" neutrino mass following the implementation of~\cite{Elbers:2024sha} and~\cite{Craig:2024tky,Green:2024xbb} but using the HiLLiPoP likelihood and also in the absence of the $z=0.7$ outliers that we have shown pull the neutrino mass bound significantly. Given our results, and if our hypotheses are correct, the preference for negative neutrino masses would vanish in that case.

From our results, we conclude that there is no significant nor compelling evidence for negative neutrino masses from cosmology. We do confirm some mild preference when Planck likelihoods affected by the lensing anomaly are adopted, pointing towards a potential residual systematic effect in these implementations. We also find that the $z=0.7$ bin of DESI, in some tension with Planck data, push towards negative values. If these outliers are confirmed, it will be interesting to see how the preference for negative masses evolves in the future. For now, and in agreement with~\cite{Allali:2024aiv}, we conclude there is no compelling evidence for negative masses from present datasets and that present constraints are still perfectly compatible with the results from neutrino oscillations.  We believe that a key check to see if cosmological data does prefer effective ``negative" neutrino masses is to do a cosmological analysis using the HiLLiPoP likelihood implementation.

%%%%%%%%%%%%%%%%%%%%%%%%%%%%%%%%%%%%%%%%%%%%%%%%%%%%%%%%%%%%%%%%%%%%%%%%%%%%
\section{Conclusions}\label{sec:conclusions}
%%%%%%%%%%%%%%%%%%%%%%%%%%%%%%%%%%%%%%%%%%%%%%%%%%%%%%%%%%%%%%%%%%%%%%%%%%%%

Recent cosmological constraints on the sum of neutrino masses ($\sum m_\nu$) already disfavour its minimum value if the mass ordering is inverted and allow very little parameter space even for normal ordering~\cite{DESI:2024mwx}. Some analyses combining additional datasets even start to rule out the minimum value of $\sum m_\nu$ for a normal ordering~\cite{Wang:2024hen}, seemingly implying tension between cosmological probes and neutrino oscillation experiments. In this context, we have critically assessed the constraints on  $\sum m_\nu$ from cosmological observables investigating their robustness against different statistical methods and determining which observables mainly dominate the budding tension with oscillations results.

To this end, we have analyzed the constraints on $\sum m_\nu$ from different datasets both through the usual Bayesian approach and through complementary frequentist methods. For the latter, given the proximity of the best fit to the physical boundary of the parameter space, we have extrapolated the likelihood profiles into the unphysical region so as to apply the appropriate correction~\cite{Feldman:1997qc}. This also allowed us to explore the preference of some datasets for negative neutrino masses recently reported in~\cite{Craig:2024tky,Allali:2024aiv,Green:2024xbb,Elbers:2024sha}. 

From the simulations and results presented in Section~\ref{sec:results} we conclude that:
\begin{itemize}

    \item \textbf{Bayesian and frequentist limits on the neutrino mass agree very well and within 10\% precision -- } 
    The agreement between the Bayesian and frequentist bounds on $\sum m_\nu$ is very good. Regarding the frequentist constraints, an extrapolation into the unphysical region of negative neutrino masses was performed to apply the appropriate correction~\cite{Feldman:1997qc} to the confidence intervals. Even though this procedure introduces a certain degree of uncertainty, we quantified the range of variation with the number of points for larger values of $\sum m_\nu$ and $\chi^2$ included and verified that the impact on the confidence intervals derived is negligible. Moreover, we also compared these with the naive cuts at the values of $\Delta \chi^2$ implied by Wilks theorem without venturing into the unphysical region and found almost identical results. When comparing these frequentist confidence intervals with the Bayesian credible regions with flat priors on $\sum m_\nu$ we find good agreement between both approaches but consistently tighter results for the frequentist method, by about $\sim 10\%$. 
    We thus conclude that the cosmological constraints on $\sum m_\nu$ are robust at this level against variations of the statistical method employed to derive them. 

     \item \textbf{The potential preference for ``negative" neutrino mass bounds is strongly dependent upon the Planck likelihood implementation used in the analysis --}  When investigating the constraints derived exclusively from Planck data, we find radically different behaviours between the Plik, CamSpec and HiLLiPoP implementations. Indeed, the preference for unphysical negative masses and hence the strongest constraints on $\sum m_\nu$ are obtained for Plik and, to a lesser degree CamSpec, and absent for HiLLiPoP, with a factor $\sim2$ weaker constrain. This tendency for weaker constraints for the CamSpec or HiLLiPoP implementations was also recently confirmed in Ref.~\cite{Allali:2024aiv}. These results seem closely correlated to the presence of the lensing anomaly in these datasets, which is reduced in the CamSpec and almost absent in the HiLLiPoP implementations. To confirm this observation, we performed analyses including the $A_\mathrm{lens}$ parameter and marginalising over its value. Interestingly, these analyses yielded the weakest constraints for the datasets more strongly affected by the anomaly. In particular, the Plik and CamSpec constraints in combination with other datasets relax by more than a factor 2. We conclude that, when the lensing anomaly is not present in the Planck likelihood either through use of the HiLLiPoP implementation or the inclusion and marginalization over $A_\mathrm{lens}$, the combined results of Planck with DESI and Pantheon relax to about $\sum m_\nu < 0.11\,{\rm eV}$ or even $\sum m_\nu < 0.2\,{\rm eV}$, both at 95\% CL, and in good agreement with both the normal and inverted orderings currently allowed by neutrino oscillation data. 

    \item \textbf{DESI-Y1 BAO measurements at $\boldsymbol{
    z= 0.7}$ pull the neutrino mass bound significantly and also induce some preference for negative values -- } When combining Planck with BAO data from DESI we observe a mild preference for unphysical negative neutrino masses, and hence a more stringent constraint. This mild preference for negative masses even appears for the HiLLiPoP implementation, for which it is absent when analyzed alone or in combination with other datasets. Upon closer examination, this preference seems to be driven mostly the BAO measurements at $z=0.7$, out of which the angular one is in $\sim \! 3\sigma$ tension with Planck expectations. Indeed, when removing this bin, the preference for negative masses disappears and the corresponding constraints relax accordingly. Similarly, if SDSS BAO data is used instead of DESI, constraints relax by about a $50\%$ in all cases considered. Thus, given the unexpectedly strong constraint derived when including the full DESI dataset, it will be very interesting to confirm if the present trend is confirmed with higher statistics from upcoming DESI data releases. 
    %Our findings indicate that a statistical fluctuation or some systematic effect with the DESI measurements at $z = 0.7$ could be behind the \EFM{unexpectedly} strong bound on the neutrino mass. \EFM{Thus, it will be very interesting to confirm if the present trend is real and gets reinforced by upcoming DESI data releases.}% \MEcom{added to please Referee:}{It may also be that this trend is real and it will be very interesting to see if it is confirmed by upcoming DESI data releases.}

    \item \textbf{The hint of dynamical dark energy relaxes the neutrino mass bound and removes the preference for ``negative'' masses -- } The results presented by the DESI collaboration~\cite{DESI:2024mwx} favour a dynamical equation of state for Dark Energy parametrized through $w_0$ and $w_a$ when extending the $\Lambda$CDM. We find that our analyses including this effect also relax significantly the constraints on $\sum m_\nu$ by a factor 2 in all cases analyzed, in agreement with the results of~\cite{DESI:2024mwx,Elbers:2024sha,Craig:2024tky}. In addition, the preference for negative neutrino mass disappears, suggesting that it may be an artifact of using the wrong cosmological model. More data are necessary to test whether the hint of dynamical dark energy, and the related artificially strong constraints to neutrino mass, are due to a statistical fluke, a systematic error, or a real deviation from $\Lambda$CDM.
    
\end{itemize}

{\bf All in all, we find it is premature to infer significant tension between present cosmological probes and neutrino oscillation data, and that the preference for unphysical negative masses is not yet compelling.} A critical test of this statement would be to have the results for the models including ``effective negative" neutrino masses using the HiLLiPoP likelihood implementation of the Planck data which does not feature a lensing anomaly. In any case, this is a very exciting period as we should be on the brink of a discovery of the absolute neutrino mass from cosmology. This would represent an indirect confirmation of the cosmic neutrino background and set a target for laboratory experiments aiming to kinematically measure the neutrino mass or, even more interestingly, its potential Majorana nature. If in the upcoming years a discovery is not made despite unprecedentedly small statistical and systematic errors, it will be a clear call for a change of paradigm, potentially signaling new physics such as neutrino decays~\cite{Escudero:2020ped,Chacko:2019nej,Escudero:2019gfk,Chacko:2020hmh,FrancoAbellan:2021hdb,Barenboim:2020vrr,Chen:2022idm}, non-standard thermodynamic histories~\cite{Farzan:2015pca,Escudero:2022gez,Alvey:2021sji,Cuoco:2005qr,Oldengott:2019lke,GAMBITCosmologyWorkgroup:2020htv}, or time varying masses~\cite{Dvali:2016uhn,Dvali:2021uvk,Lorenz:2018fzb,Lorenz:2021alz,Esteban:2021ozz,Esteban:2022rjk,Sen:2023uga,Sen:2024pgb}. It is clear that upcoming years are of utmost importance for neutrino cosmology, as whether neutrino masses are detected or not, we are on the verge of a major breakthrough.

%%%%%%%%%%%%%%%%%%%%%%%%%%%%%%%%%%%%%%%%%%
\medskip
\paragraph*{Acknowledgments.}
EFM acknowledges very interesting discussions with Stefano Gariazzo and thanks him for pointing out relevant references. We also warmly thank Thomas Schwetz for very interesting comments and suggestions for our analysis. This project has received support from the European Union’s Horizon 2020 research and innovation programme under the Marie Skłodowska-Curie grant agreement No~860881-HIDDeN and No 101086085 - ASYMMETRY, and from the Spanish Research Agency (Agencia Estatal de Investigaci\'on) through the Grant IFT Centro de Excelencia Severo Ochoa No CEX2020-001007-S and Grant PID2022
137127NB-I00 funded by MCIN/AEI/10.13039/501100011033. 
We acknowledge support from the HPC-Hydra cluster at IFT and the HTCondor cluster at CERN.
XM acknowledges funding from the European Union’s Horizon Europe Programme under the Marie Skłodowska-Curie grant agreement no.~101066105-PheNUmenal. The work of DNT was supported by the
Spanish MIU through the National Program FPU (grant number FPU20/05333).

%TO DO LIST:
%\begin{enumerate}
%    \item Quantify a bit better the story about the 2 and 4pt lensing correlation that~\cite{Green:2024xbb} et al find. 
%    \item Run the Bayesian and frequentist analyses for Hillipop+DESIno07 to add to the text.
%    \item Do the full Planck runs but including the lensing likelihoods.
%    \item Fig.~\ref{fig:planck_MAX} maybe change colors. I think it may be a bit confusing to use the same ones. 
%\end{enumerate}

\bibliography{biblio}

\newpage
\onecolumngrid

\appendix

%%%%%%%%%%%%%%%%%%%%%%%%%%%%%%%%%%%%%%%%%%%%%%%%%%%%%%%%%%%%%%%%%%%%%%%%%%%%
\section{Posterior distributions}
\label{app:posteriors}
%%%%%%%%%%%%%%%%%%%%%%%%%%%%%%%%%%%%%%%%%%%%%%%%%%%%%%%%%%%%%%%%%%%%%%%%%%%%
In this Appendix we draw the $\sum m_\nu$ posterior distributions for some of our analyses. In particular, we present in Fig.~\ref{fig:planck_only_posterior} the Bayesian analogue of Fig.~\ref{fig:mnu_planck_only}. Secondly, the Bayesian analyses for the different Planck+BAO combinations are shown in Fig.~\ref{fig:planck_bao_posterior} (see Fig.~\ref{fig:planck_BAO} for their frequentist counterpart). Lastly, the Planck+DESI-Y1+Pantheon+ posteriors are presented in Fig.~\ref{fig:planck_max_posterior}, where we are using the same datasets as in the frequentist analyses of Fig.~\ref{fig:planck_MAX}.

For the comparison between the bounds extracted from the Bayesian and frequentist approaches, we refer the reader to Tables~\ref{tab:Planckonly},~\ref{tab:PlanckBAO} and~\ref{tab:PlanckBAOSN}.

\begin{figure*}[h!]
    \centering
    \includegraphics[width=0.45\textwidth]{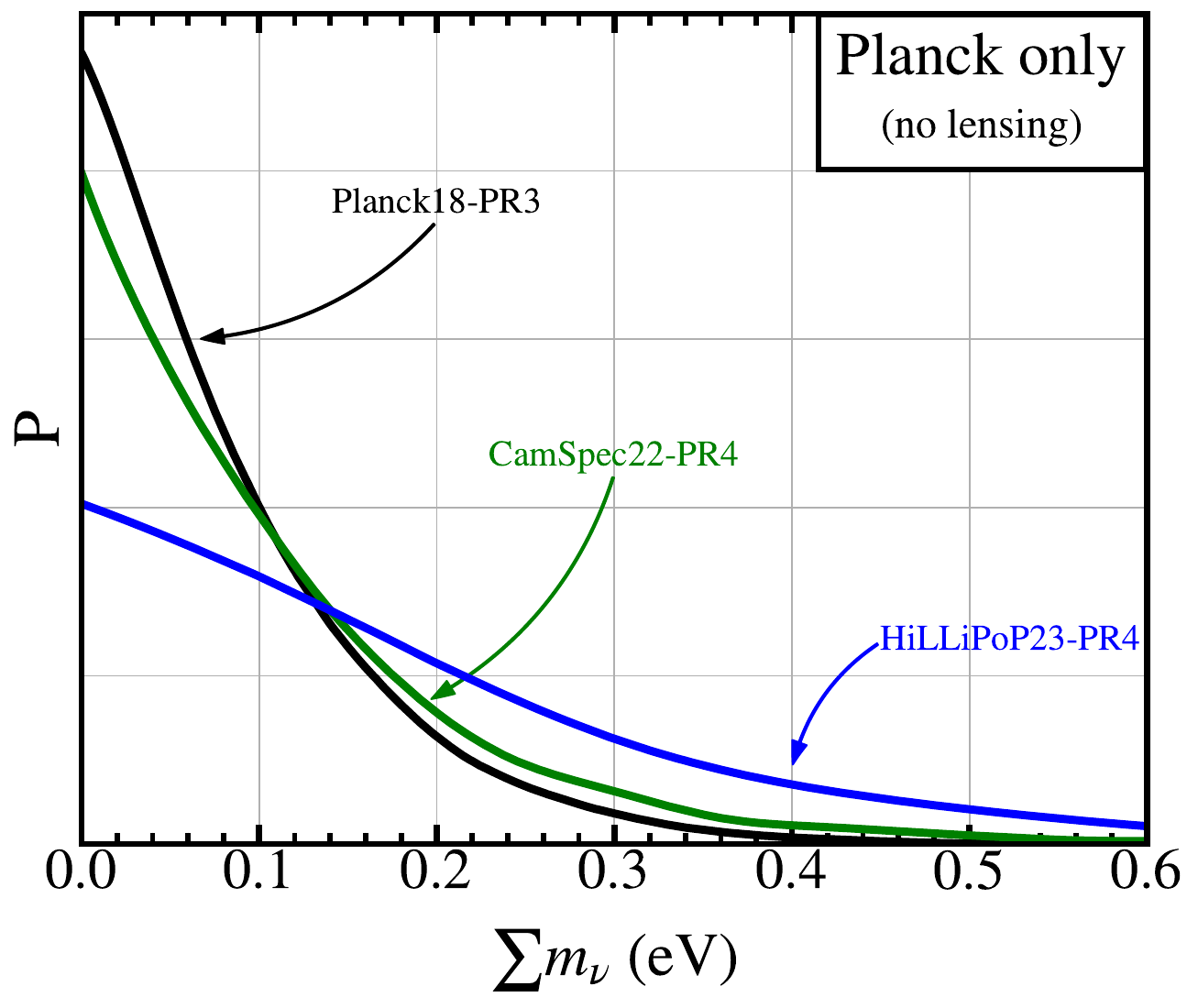}
    \caption{Posterior distributions for the same data set combinations as in Fig.~\ref{fig:mnu_planck_only} but analyzed within a Bayesian framework.}
    \label{fig:planck_only_posterior}
\end{figure*}

\begin{figure*}[h!]
    \centering
    \includegraphics[width=0.45\textwidth]{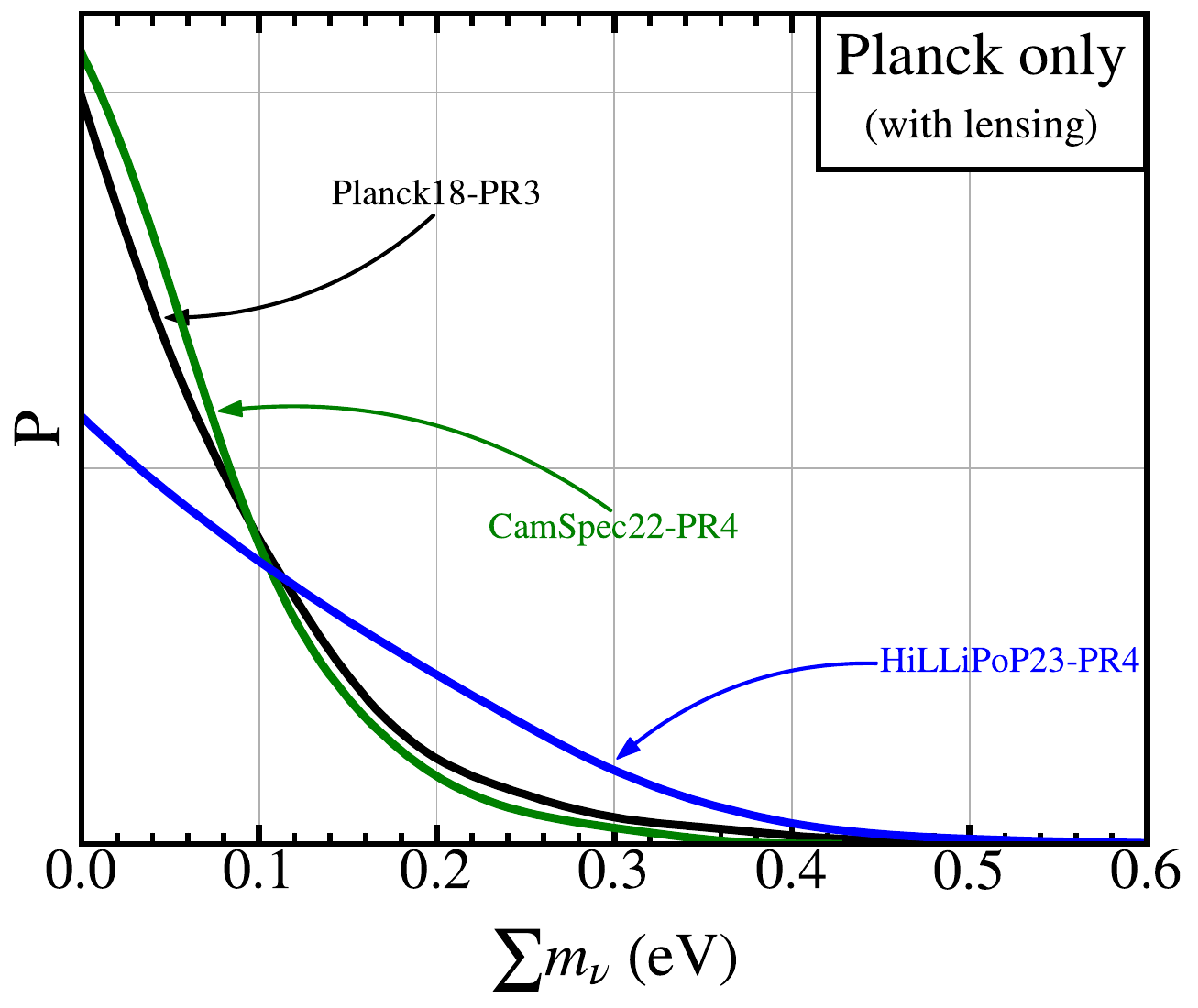}
    \caption{Posterior distributions for the same data set combinations as in Fig.~\ref{fig:mnu_planck_lensing} but analyzed within a Bayesian framework.}
    \label{fig:planck_lens_posterior}
\end{figure*}

\begin{figure*}[h!]
    \centering
    \includegraphics[width=0.9\textwidth]{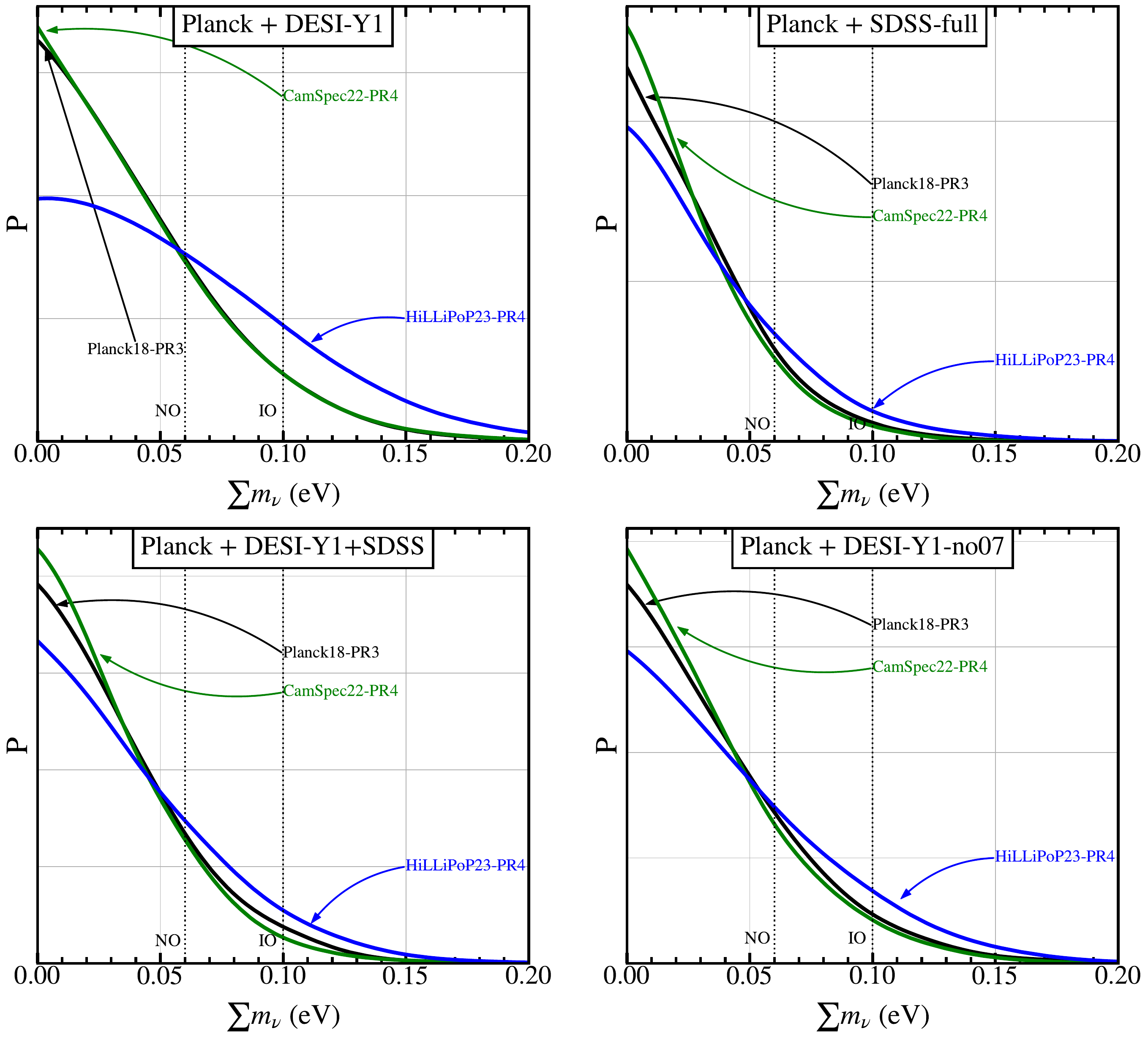}
    \caption{Posterior distributions for the same data set combinations as in Fig.~\ref{fig:planck_BAO} but analyzed within a Bayesian framework.}
    \label{fig:planck_bao_posterior}
\end{figure*}

\begin{figure*}[h!]
    \centering
    \includegraphics[width=0.95\textwidth]{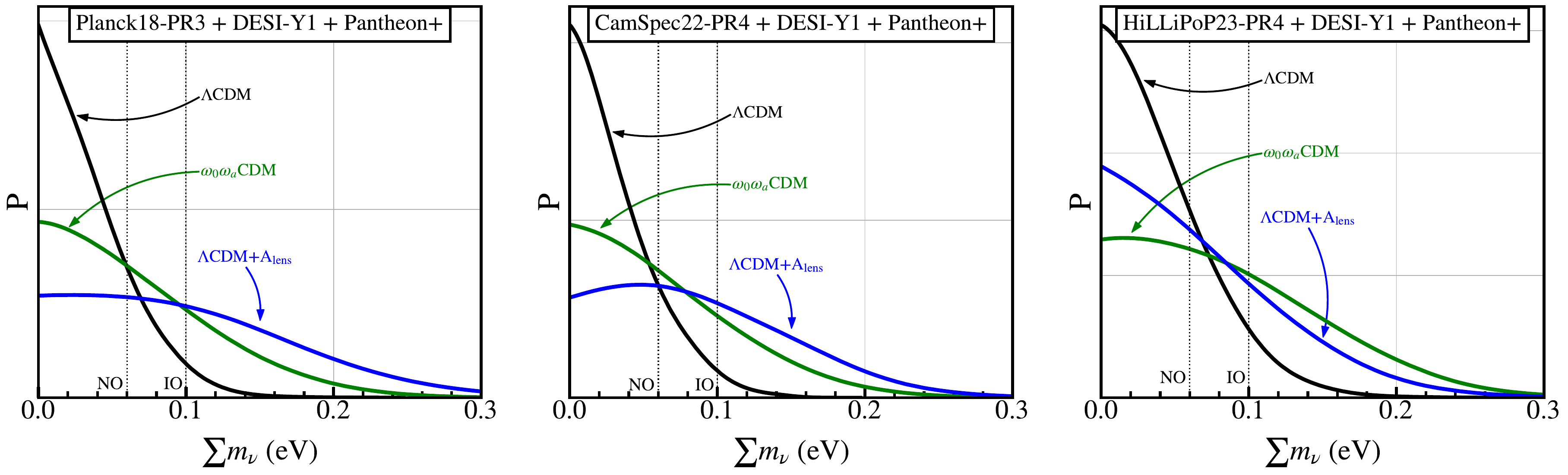}
    \caption{Posterior distributions for the same data set combinations as in Fig.~\ref{fig:planck_MAX} but analyzed within a Bayesian framework.}
    \label{fig:planck_max_posterior}
\end{figure*}

%%%%%%%%%%%%%%%%%%%%%%%%%%%%%%%%%%%%%%%%%%%%%%%%%%%%%%%%%%%%%%%%%%%%%%%%%%%%
\section{Comparison with other Gaussian extrapolations}\label{app:comparison}
%%%%%%%%%%%%%%%%%%%%%%%%%%%%%%%%%%%%%%%%%%%%%%%%%%%%%%%%%%%%%%%%%%%%%%%%%%%%

For the sake of assessing the robustness of our results when performing the extrapolation of the likelihood profiles into the unphysical $\sum m_\nu$ region, we compare our results with those of other works, such as Refs.~\cite{Allali:2024aiv,Elbers:2024sha}. In these works, a similar procedure was performed but within a bayesian framework: the authors fitted to a Gaussian their posterior in the physical region and extrapolated to negative masses.

The comparison is shown in Fig.~\ref{fig:planck_bao_extrapolation_comparison}, where we plot our $\Delta\chi^2$ profiles and their $-2\log \mathcal{P}$ for similar datasets. We find an overall agreement in the preference for a best fit in the negative neutrino mass region, suggesting that the conclusions drawn from the extrapolation are not largely dependent on the statistical procedure employed.
\begin{figure*}[h!]
    \centering
    \includegraphics[width=0.82\textwidth]{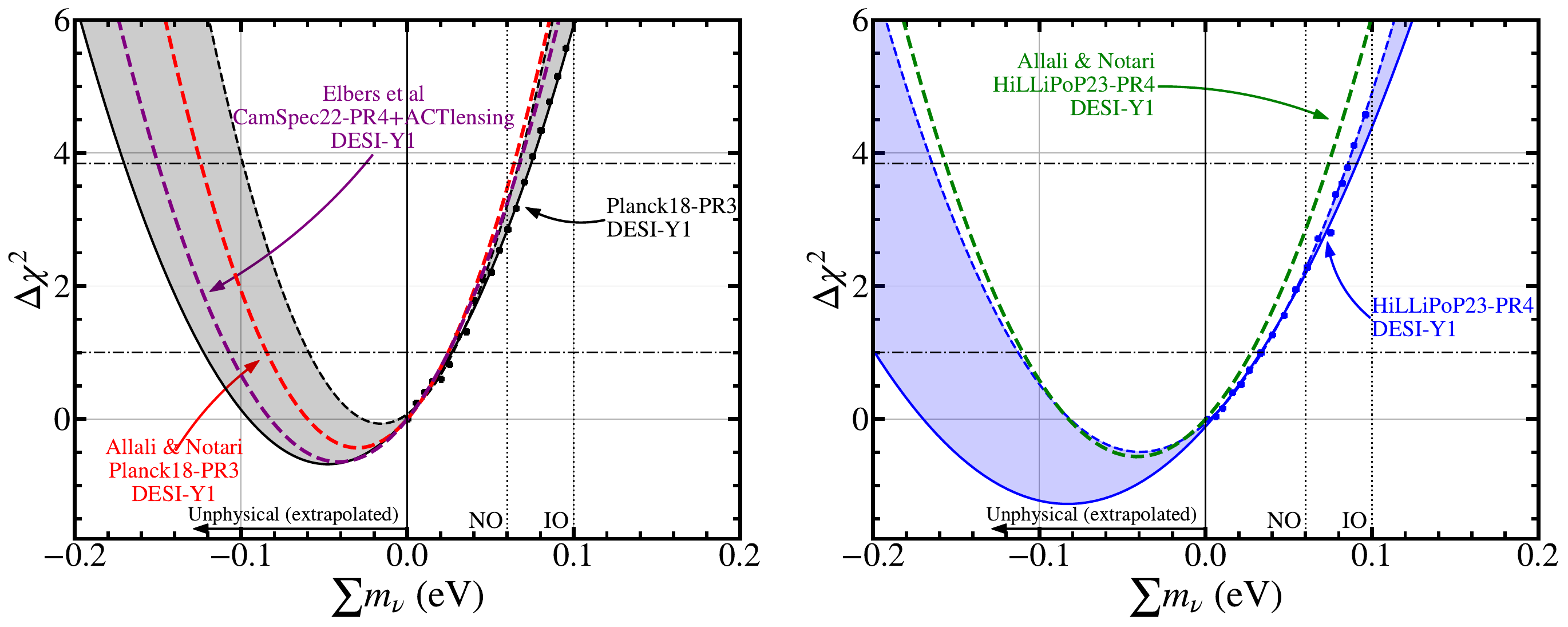}
    \caption{Comparison between our $\Delta\chi^2$ profiles and their gaussian extrapolation with the corresponding extrapolated $-2\log \mathcal{P}$ of Ref.~\cite{Allali:2024aiv} and Ref.~\cite{Elbers:2024sha}.}
    \label{fig:planck_bao_extrapolation_comparison}
\end{figure*}

%\newpage
%%%%%%%%%%%%%%%%%%%%%%%%%%%%%%%%%%%%%%%%%%%%%%%%%%%%%%%%%%%%%%%%%%%%%%%%%%%%
\section{Correlations with other parameters}\label{app:correlations}
%%%%%%%%%%%%%%%%%%%%%%%%%%%%%%%%%%%%%%%%%%%%%%%%%%%%%%%%%%%%%%%%%%%%%%%%%%%%

It has been shown that extending the standard cosmological model either with the A$_{\text{lens}}$ parameter or by adding a dynamical dark energy equation of state $w(a)=w_0+(1-a)w_a$, the cosmological bound on neutrino masses can be substantially relaxed and thus reduce the tension with oscillation data. In Fig.~\ref{fig:planck_mnu_w0_correlation} we present the correlations of $\sum m_\nu$ with $w_0$ and A$_{\text{lens}}$, as derived from our frequentist and Bayesian analyses by plotting the $w_0$ (left panel) and $A_{\text{lens}}$ (right panel) values of the profile likelihood points as well as the Bayesian contours. Most interestingly, as previously stated, \texttt{HiLLiPoP23-PR4} yields the closest A$_{\text{lens}}$ value to $\Lambda$CDM.

\begin{figure*}[h!]
    \centering
    \includegraphics[width=0.82\textwidth]{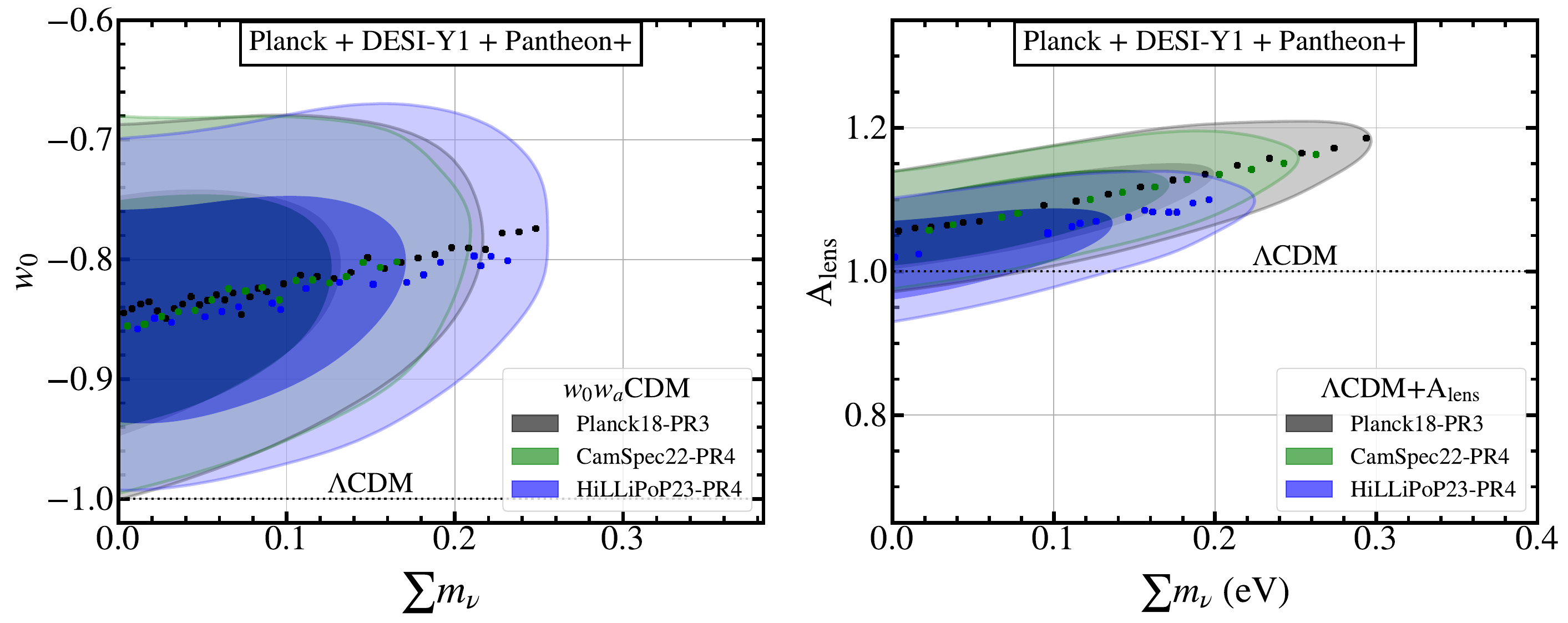}
    \caption{Correlations of $\sum m_\nu$ with $w_0$ and A$_{\text{lens}}$. The points are extracted from our profile likelihoods for the Planck+DESI-Y1+Pantheon+ combination, while the shaded regions correspond to the bayesian credible intervals at 1 and 2$\sigma$.}
    \label{fig:planck_mnu_w0_correlation}
\end{figure*}

\end{document}